\documentclass[traditabstract]{aa} 

\usepackage{aalongtable}
\usepackage{supertabular}
\usepackage{graphicx}
\usepackage{txfonts}

\newcommand{\hi}{\mbox{H\,{\sc i}}} 
\newcommand{\hii}{\mbox{H\,{\sc ii}}}

\newcommand{\feii}{\mbox{Fe\,{\sc ii}}} 
\newcommand{\mgii}{\mbox{Mg\,{\sc ii}}} 
\newcommand{\siii}{\mbox{Si\,{\sc ii}}}

\newcommand{\suii}{\mbox{S\,{\sc ii}}}

\newcommand{\znii}{\mbox{Zn\,{\sc ii}}}

\newcommand{\niii}{\mbox{N\,{\sc iii}}}

\begin{document}

     \title{The cosmic evolution of dust-corrected metallicity\\in the neutral gas\thanks{Based on observations carried out at the European Organisation for Astronomical Research in the Southern Hemisphere under ESO programmes 065.P-0038, 065.O-0063, 066.A-0624, 067.A-0078, and 068.A-0600.}}

   \author{Annalisa De Cia
          \inst{1}
          \and
          C\'edric Ledoux \inst{2} \and Patrick Petitjean \inst{3} \and Sandra Savaglio \inst{4}
          }

\institute{
European Southern Observatory, Karl-Schwarzschild Str. 2, 85748 Garching bei M\"unchen, Germany; \email{adecia@eso.org}
\and
European Southern Observatory, Alonso de C\'ordova 3107, Casilla 19001, Vitacura, Santiago 19, Chile
\and 
IAP, CNRS and Universit\'e Paris 6, 98bis Boulevard Arago, 75014 Paris, France
\and
Physics Dept., University of Calabria, via P. Bucci, 87036, Arcavacata di Rende, Italy
}

   \date{Received September 19, 2017; accepted Month dd, 2017}

   \abstract
{
Interpreting abundances of Damped Ly-$\alpha$ Absorbers (DLAs) from absorption-line spectroscopy has typically been a challenge because of the presence of dust. Nevertheless, because DLAs trace distant gas-rich galaxies regardless of their luminosity, they provide an attractive way of measuring the evolution of the metallicity of the neutral gas with cosmic time. This has been done extensively so far, but typically not taking proper dust corrections into account. The aims of this paper are to: \textit{i)} provide a simplified way of calculating dust corrections, based on a single observed [$X$/Fe], \textit{ii)} assess the importance of dust corrections for DLA metallicities and their evolution, and \textit{iii)} investigate the cosmic evolution of iron for a large DLA sample. We have derived dust corrections based on the observed [Zn/Fe], [Si/Fe], or [S/Fe], and confirmed their robustness. We present dust-corrected metallicities in a scale of [Fe/H]$_{\rm tot}$ for 236 DLAs over a broad range of $z$, and assess the extent of dust corrections for different metals at different metallicities. Dust corrections in DLAs are important even for Zn (typically of 0.1--0.2, and up to $0.5$~dex), which is often neglected. Finally, we study the evolution of the dust-corrected metallicity with $z$. The DLA metallicities decrease with redshift, by a factor of 50--100 from today to $\sim12.6$ billion years ago ($z=5$). When including dust corrections, the average DLA metallicities are 0.4--0.5~dex higher than without corrections. The upper envelope of the relation between metallicity and $z$ reaches solar metallicity at $z\lesssim0.5$, although some systems can have solar metallicity already out to $z\sim3$.
}


   \maketitle

\section{Introduction}

In the past 50 years, the launch of satellites with spectroscopic UV capability has opened up the possibility of characterizing the metal abundances in the Galactic neutral Interstellar Medium \citep[ISM, that is, dominated by \hi{} and singly ionized metals, e.g.,] []{Draine11b}. A large number of studies \citep{Jenkins73,Field74,Morton75,Cardelli93,Hobbs93,Phillips82,Phillips84,Jenkins86,Savage91,Savage92,Welty95,Savage96} have taken the opportunity to quantify the metal abundances of elements $X$ in the Galactic ISM, [$X$/H] $\equiv \log( N(X)/ N({\rm H})) - \log( N(X)_\odot /N({\rm H})_\odot )$. The observed abundances show large variations, which depend on the refractory properties of the observed metals. The higher the condensation temperature of a metal (as measured in the lab), the lower its abundance that we can measure in the gas-phase \citep[e.g.,] []{Field74,Jenkins86,Cardelli93, Hobbs93,Phillips82,Phillips84,Welty95,Savage96}. It became soon evident that large fractions of the refractory metals were missing from the observed gas-phase, because they were instead locked into dust grains. This phenomenon is called dust depletion. A jump forward in the understanding of dust depletion was made by \citet{Jenkins09}, who discovered that the metal abundances in the ISM in the Galaxy correlate with each other. He found compelling dust depletion sequences for several metals, smoothly evolving from less to more dusty clouds in the Galactic ISM. However, until recently the study of dust depletion in the Galaxy was done by assuming that the underlying, intrinsic metallicity of the ISM is solar. Any deviation from solar abundances had been attributed to dust depletion. 

\citet{DeCia16}, hereafter Paper I, studied the abundances of the Galactic ISM, using data from \citet{Jenkins09}, and Damped Lyman-$\alpha$ Absorbers (DLAs) at $2 \leq z \leq 4$ from the sample selected by \citet{Ledoux06}, and without making any assumption on the intrinsic metallicity. DLAs are typically subsolar-metallicity systems characterized in absorption toward distant Quasars. They are associated with gas in and around low-mass galaxies \citep[e.g.,] []{Christensen14}, and are the largest reservoirs of neutral hydrogen in the Universe \citep[e.g.,] []{Wolfe95}. The large neutral hydrogen columns of DLAs \citep[defined as $N(\mbox{\hi})\geq20.3$,] []{Wolfe05} shield the gas against ionization \citep[which is otherwise crucial for systems with lower $N(\mbox{\hi})$,][]{Viegas95,Vladilo01,Peroux07,Ledoux09,Milutinovic10,DeCia11,DeCia12,Vreeswijk13}. Thus, the metallicity measurements do not depend on difficult ionization corrections. In addition, the existence of the same correlations of relative abundances (dust depletion sequences) in different environments such as the Milky Way, the Magellanic Clouds, and DLAs suggests that DLAs behave like ISM gas, where dust grains can grow in a similar way (De Cia et al. in prep.). Gas from DLA can in fact include different ISM-like phases such as the warm neutral medium, the cold neutral medium, and the diffuse molecular gas \citep[e.g.,] []{Draine11b}, it can extend well outside the stellar disks as extra-planar gas, but does not have the full extent of the halo \citep{Pontzen08,Marasco11,Lehner15,Tumlinson17}. Local gas-rich dwarf galaxies, which may resemble part of the DLA population at different $z$, are gas dominated (gas-to-baryonic fractions of 50--90\%) and their main \hi{} disks are more extended than their stellar disks by a factor of four on average \citep{Lelli16}.

Because DLAs trace gas-rich galaxies out to high redshift, and regardless of their luminosity, they provide an attractive way of measuring the evolution of the metallicity of the neutral gas with cosmic time. This has been studied extensively \citep{Prochaska03,Kulkarni07,Dessauges-Zavadsky08,Rafelski12}, but without taking the effect of dust into account. Several studies have recognized that dust depletion is an important phenomenon for DLAs too \citep[e.g.,] []{Pettini94,Kulkarni97,Vladilo98,Pettini00,Savaglio01,Ledoux02,Prochaska02,Vladilo02,Dessauges-Zavadsky06,Rodriguez06,Meiring06,Vladilo11,Som13,DeCia13,Quiret16,Wiseman17}. \citet{Calura03} and \citet{Lanfranchi03} have studied the chemical evolution in DLAs taking some correction for dust depletion into account, based on the dust-correction models of \citet{Vladilo02}. \citet{Savaglio06} presented simplified rough dust corrections, sometimes even based on the column density of an individual metal, such as iron. 
We developed a method to characterize dust depletion without any assumption on the gas metallicity. This was achieved through the study of relative abundances, $[X/Y] = \log( N(X)/ N(Y)) - \log( N(X)_\odot /N(Y)_\odot ) $, of several metals with different nucleosynthetic and refractory properties. The amount of dust in a system can be traced by the observed [Zn/Fe], because Fe is much more depleted than Zn, and the two elements tend to follow each other nucleosynthetically, at least in the metallicity range of interest here ($-2\leq [M/{\rm H}]\leq 0$, see Sect. \ref{sect bias}). Other ratios, such as [Ti/Si] or [Si/S] can be also used (see correlations with [Zn/Fe], Paper I), but are typically less constrained. In Paper I, we discovered tight correlations between relative abundances, for DLAs as well as for the Galaxy ISM. This way we could determine the dust depletion in each system, based on the observed relative abundances, and in particular [Zn/Fe], and without assumptions on the reference metallicity of the gas. With this method, it is now possible to first determine the dust depletion from the relative abundances, and then correct for dust depletion and derive the dust-corrected, total gas metallicity, in a fairly accurate way.

In this paper we develop and provide a way of calculating dust-corrected metallicities (Sect. \ref{sec dust corrections}), which is simplified with respect to Paper I and is based on an observed [$X$/Fe]. We then apply this method to the largest possible DLA sample with [$X$/Fe] measurements and a broad range in redshift $z$ (Sect. \ref{sect sample}). In Sect. \ref{sec importance dust corr} we assess the importance of dust corrections in the DLA population. Finally, we show the evolution of the neutral gas metallicity with cosmic time (Sect. \ref{sec met evolution}), taking dust corrections into account. These corrections, as we discuss below, are crucial for metal-rich systems. We discuss potential selection effects and caveats in Sect. \ref{sect bias}, and summarize our results and conclusions in Sect. \ref{sect conclusions}.

\section{Dust corrections}
\label{sec dust corrections}
In Paper I, we found robust dust-corrected metallicities based on the simultaneous study of the abundances of several metals and on the dust depletion sequences, which were globally discovered for the whole sample of DLAs and Galactic clouds. Such dust corrections, based on the observations of several metals, are the most reliable. 

However, often only a limited subset of metals can be constrained. Therefore we have derived and tested here a simplified method to calculate dust corrections, based on a single abundance relative to Fe, [$X$/Fe]. We call this method the single-reference method.

\subsection{Based on [Zn/Fe]}

When available, this method uses [Zn/Fe] to estimate the dust depletion given the depletion sequences of Paper I. The first step is to calculate the depletion of an element $X$ as follows:
 \begin{equation}
\delta_X =  A2_{X} + B2_{X}\times {\rm [Zn/Fe]}\mbox{,}
\label{eq delta}
\end{equation}
where the coefficients for Fe are $A2_{{\rm Fe}}=-0.01\pm0.03$ and $B2_{{\rm Fe}}=-1.26\pm0.04$, and the full list of coefficients for each metal is reported in Table 3 of Paper I. The dust-corrected metallicity reference is then:
\begin{equation}
[{\rm Fe/ H}]_{\rm tot}=[{\rm Fe/H}] - \delta_{\rm Fe} \mbox{,}
\label{eq FeH_tot}
\end{equation}
where [Fe/H] are the observed abundances, and $\delta_{\rm Fe}$ is calculated with Eq.\ref{eq delta}. Here we chose [Fe/H]$_{\rm tot}$ as a reference for metallicity for two reasons. First, Fe is not affected by $\alpha$-element enhancement [$\alpha$/Fe], and therefore no nucleosynthetic corrections are needed in Eq. \ref{eq FeH_tot}. Second, Fe is among the metals that are most easily measured, and therefore most widely available. In addition, $[{\rm Fe/ H}]_{\rm tot}$ is widely used as a reference for stellar abundances, which can be an useful comparison. We estimated the uncertainty on $[{\rm Fe/ H}]_{\rm tot}$ as the quadratic sum of the uncertainty on $[{\rm X/Fe}]$, $N$(\hi{}), and some allowance for the uncertainty on the slope parameter $B1_{X}$, which we assumed to be 0.07~dex (from Paper I).

The coefficients for the depletion sequences in Eq. \ref{eq delta} were calculated in Paper I making a small assumption on the distribution of the $\alpha$-element enhancement with metallicity, namely that the $\alpha$-element knee in DLAs is located at the same metallicity as in the Milky Way. This is not necessarily true because local low-mass galaxies show an $\alpha$-element knee at lower metallicities \citep{deBoer14}. Nevertheless, this assumption has a minimal effect on the calculation of the dust-depletion sequences.

\subsection{Based on [Si/Fe] or [S/Fe]}

When [Zn/Fe] is not directly observed, then the relative abundance [Si/Fe] or [S/Fe] can be used instead. The depletion sequences are defined as a function of [Zn/Fe] (Eq. \ref{eq delta}). However, the relative abundances [$X$/Zn] correlate empirically with [Zn/Fe] 
(Eq. 1 and Fig. 3 of Paper I), and therefore we derived the expected [Zn/Fe]$_{\rm exp}$ from the observed [$X$/Fe] as follows: 
\begin{equation}
\begin{array}{lcl}
[X / {\rm Fe}] &=& [X/{\rm Zn}] + [{\rm Zn/Fe}]_{\rm exp} \\
               &=& A1_{X} + B1_{X}\times {\rm [Zn/Fe]} + [{\rm Zn/Fe}]_{\rm exp}\\
\end{array}   
\label{eq XFe ZnFe 1}  
\end{equation}
\begin{equation}
\begin{array}{lcl}         
[{\rm Zn/Fe}]_{\rm exp} &=& \left(  [X /{\rm Fe}] - A1_{X}  \right) / (B1_{X} + 1)\mbox{,}
\end{array}     
\label{eq XFe ZnFe 2}
\end{equation}
where [$X$/Fe] are the observed abundances, and we equated [Zn/Fe] to [Zn/Fe]$_{\rm exp}$ to derive Eq. \ref{eq XFe ZnFe 2} from Eq. \ref{eq XFe ZnFe 1}. The coefficients for Si (S) are $A1_{{\rm Si}}=0.26\pm0.03$ and $B1_{{\rm Si}}=-0.51\pm0.06$ ($A1_{{\rm S}}=0.25\pm0.03$ and $B1_{{\rm S}}=-0.23\pm0.07$). The full list of coefficients for each metal is reported in Table 2 of Paper I. The coefficient used in Eq. \ref{eq XFe ZnFe 1} were measured in Paper I on a purely empirical basis, and are not subject to any assumptions. Indeed, the empirical correlations between [$X$/Zn] and [Zn/Fe] are observed for several metals, including $\alpha$-elements. This means that possible enhancement of Si and S will be already naturally accounted for in these empirical correlations, and thus it is not necessary to make any assumption on $\alpha$-element enhancement.

\subsection{Robustness of the method}
\begin{figure}
   \centering
   \includegraphics[width=9cm]{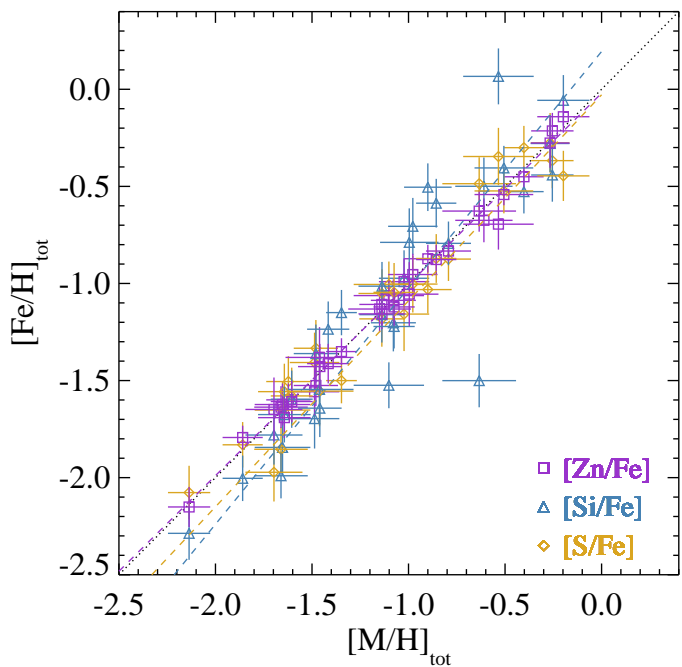}
   \caption{Comparison between the solid dust-corrected metallicities [$M$/H]$_{\rm tot}$ derived in Paper I from several metals simultaneously and the dust-corrected metallicities [Fe/H]$_{\rm tot}$ derived with the single-reference method (Sect. \ref{sec dust corrections}). The linear fits to the data (dashed curves) have slopes of 0.98, 1.22, and 1.06 for Zn, Si, and S, respectively, and intercepts of $-0.02$, $0.19$, and $-0.03$. The differences from the one-to-one line are taken into account in the paper for the final calculation of the [Fe/H]$_{\rm tot}$.}
              \label{fig test met}%
    \end{figure}

We tested the reliability of the single-reference method on the DLA sample of Paper I, for which we have robust dust-corrected metallicity [$M$/H]$_{\rm tot}$ based on the observations of several metals. In Fig. \ref{fig test met} we show a comparison between the robust dust-corrected metallicities [$M$/H]$_{\rm tot}$ and the dust-corrected metallicities derived with the single-reference method based on the observed [Zn/Fe] (squares), [Si/Fe] (triangles), and [S/Fe] (diamonds). The dust-corrected metallicities calculated using the observed [Zn/Fe] are the most reliable, among the single-reference metallicities, and almost perfectly trace the real dust-corrected metallicities. The [Si/Fe] and [S/Fe] single-reference metallicities overall follow the real dust-corrected metallicities. However, for [Si/Fe], the single-reference metallicities are a bit underestimated at low metallicity (by $-0.25$~dex at [$M$/H]$_{\rm tot}=-2$) and a bit overestimated at high metallicity (by $0.2$~dex at [$M$/H]$_{\rm tot}=0$). This is visible from the fit to the data in Fig. \ref{fig test met}. To account for this effect and avoid related biases, we further corrected the [Fe/H]$_{\rm tot}$ that we derive with the single reference method by the difference between the fit in Fig. \ref{fig test met} and the one-to-one line, which varies a bit with metallicity. While for Zn-based measurements this is negligible, this affects mostly the Si-based measurements, by an average of 0.18~dex, with a standard deviation of 0.15~dex. The main effect of this additional correction is at high-$z$, where Si-based measurements are dominant, and it is most notable for systems with very low intrinsic metallicity, the most extreme cases being a correction of 0.44~dex at a [Fe/H]$_{\rm tot}=-2.95$ which was corrected to [Fe/H]$_{\rm tot}=-2.51$ to account for the Si trend in Fig. \ref{fig test met}. The potential need for an update of the depletion-sequences coefficients for Si is discussed in the Appendix.

While the [S/Fe] dust corrections seem more accurate than those based on [Si/Fe], we caution that S may be a troublesome element. Indeed, the estimates of [S/H] toward stars have been problematic for the Galaxy \citep[e.g.,] []{Jenkins09} and the Small Magellanic Cloud \citep[e.g.,] []{Jenkins17}, potentially because of a ionization problem. The lines of sight in these environments may cross \hii{} regions where H may be ionized, while S may be mostly in \suii. This is less likely to be the case for DLAs, where lines of sight are less likely to penetrate such \hii{} regions. The depletion sequences observed in Paper I for S indicate that DLAs do not seem to suffer from the \suii{} ionization problem. However, more DLA observations at high [Zn/Fe] are needed for this matter to be finally settled. Until then, we still recommend exerting some caution in using [S/Fe] as a reference for dust depletion. Thus, we derived dust corrections with the single-reference method using [Zn/Fe] when available, otherwise [Si/Fe], or otherwise [S/Fe].

We note that in Fig. \ref{fig test met} there are two strong outliers in the relations, all associated with Si measurements with fairly large errors, namely Q~1444-014 and Q~2359-022\footnote{Q~1444-014 and Q~2359-022 are presented in \citet{Ledoux03} and Paper I, respectively. We did not find any obvious problem with their estimation of the \siii{} column densities.}. This has no tangible effect for this work, because we aim at characterizing the whole population and such outliers are statistically not important. However, possible fluctuations should be taken into account when using the single-reference method based on [Si/Fe] to correct for dust depletion in a given individual system.

The potential non-accuracy of the column density measurements may introduce some bias, in particular for the large sample (see Sect. \ref{sect sample}). Indeed, in 47 cases the [Zn/Fe]$_{\rm exp}$ is negative in the large sample\footnote{down to $-1.5$~dex, with a mean and standard deviation of $-0.3$ and $0.4$, respectively. See Table C.2 for the individual values.}, resulting in a non-physical positive depletion, for which we assigned $\delta_X\equiv0$ (that is, no dust correction needed). However, only three out of these systems have actual Zn measurements ([Zn/Fe] $=-0.15,-0.32,-0.02$~dex), the others are derived from either [Si/Fe] or [S/Fe]. As a comparison, for the clean sample the [Zn/Fe]$_{\rm exp}$ is negative only in two cases, and only by a small amount (that is, $-0.08$, $-0.19$~dex). Nucleosynthetic effects on [Zn/Fe] should be small, as discussed in Sect. \ref{sec znfe}. Therefore the 47 negative [Zn/Fe]$_{\rm exp}$ in the large sample are likely the product of inaccurate estimates of the column densities. While a complete inspection of all cases is out of the scope of this paper, we selected all the DLAs with [Zn/Fe]$_{\rm exp}\leq -0.3$ and with a small quoted uncertainty (error on the observed [X/Fe] $< 0.1$~dex) for a close inspection. All these cases (11) revealed problems, for example saturation of the \siii{} lines used is underestimated, especially in low-resolution data. The details of these measurements are reported in the Appendix.

\subsection{The reliability of [Zn/Fe] as a dust indicator}
\label{sec znfe}

In this paper we derived dust corrections based on the observed or expected [Zn/Fe]. The underlying assumption is that Zn and Fe trace each other nucleosynthetically. Zn is not strictly an iron-peak element, and it is produced in both core-collapse and Type Ia SNe \citep[e.g.,] []{Nomoto97}, in quantities that depend on the adopted SN model. A flat [Zn/Fe] (and slightly supersolar) is observed in the Galactic stellar populations for $-2\leq [M/{\rm H}]\leq 0$ \citep[e.g.,] [using LTE estimates]{Sneden91,Saito09}. There are strong ($+0.5$~dex) deviations from this at lower ($[M/{\rm H}]<-3$) and higher metallicities \citep[e.g.,] []{Primas00,Nissen07}. In particular, \citet{Nissen07} showed that, when non-LTE effects in Galactic stellar abundances are taken into account, Zn behaves somewhat like an $\alpha$-element, but with a very small [Zn/Fe] amplitude (0.1--0.2~ dex) and an $\alpha$-element "knee" at low metallicities ($\sim1$ dex lower than for the other $\alpha$-elements). \citet{Nissen11} found two distinct stellar populations in the Galactic halo, one with nearly solar [Zn/Fe] values, and one with higher $\alpha$-element enhancement and a near-constant [Zn/Fe] $\sim 0.15$. Therefore Zn can be consider as a hybrid element. Overall, different SNe types and stellar populations contribute with time to the [Zn/Fe] observed in the ISM, and this does not show strong deviations from the solar value. Indeed, the observed [Zn/Fe] values in DLAs converge toward zero for low metallicities (below [M/H] $\sim -2$, Paper I). In this regime the effects of dust depletion is negligible, and we do not observe an $\alpha$-element plateau of positive [Zn/Fe]. In addition, the mere existence of narrow dust depletion sequences for the Galaxy as well as DLAs (Paper I) limits the possible scatter in the nucleosynthetic [Zn/Fe] of the current sample to be $<0.2$~dex. While small deviations from a solar [Zn/Fe] are possible, as mentioned above, we exclude large nucleosynthetic effects on the [Zn/Fe] abundances. Indeed, we do not measure heavily negative [Zn/Fe] in DLAs. The use of [Zn/Fe] as a dust tracer in the ISM is discussed also in Appendix A of Paper I.

Stellar measurements of [Zn/Fe] can vary dramatically, in gas-poor environments such as dwarf spheroidals \citep{Skuladottir17} and the inner bulge of the Milky Way \citep{Barbuy15,Duffau17}. However, gas-rich environment, such as the Milky Way disk, the Magellanic Clouds, and DLAs, do not show this effect in their gas component (e.g., De Cia et al. in prep.), where the contribution from different stellar populations have slowly been reprocessed in the gas. It is possible that the [Zn/Fe] zeropoint is slightly off, perhaps down to $-0.2$~dex, as discussed in Paper I. The truncation in the distribution of the $\delta_{\rm Zn}$ (Fig. \ref{fig hist delta}) may also support the possibility of such small offset. In the [Zn/Fe] $=-0.2$~dex offset case, Si would be depleted by 0.1~dex more heavily than with the current assumption, and much less than this for S and Zn. This would slightly push up the metallicity, by less than 0.1~dex, at the high-$z$ end.

\section{Data}
\label{sect sample}

We calculated dust-corrected metallicities [Fe/H]$_{\rm tot}$ for DLAs taken from different samples. First, we included the DLAs from Paper I, which have robust dust-corrected metallicities derived from studying simultaneously all metals. Then we selected the DLAs using two approaches, which we called the clean and the large samples, as explained below.

\subsection{The clean sample}
We selected DLAs that have high-quality data and which abundances can be trusted, in the sense that are all based on high-resolution spectroscopy and we have checked them thoughroughly. For this we resourceed to the ``quality'' selection of \citet{Moller13}, which includes data from \citet{Ellison12}, \citet{Peroux06}, \citet{Pettini00}, \citet{Peroux08}, \citet{Rao05}, \cite{Meiring11}, \citet{Pettini99} at $z<2$, and \citet{Ledoux06} and \citet{Rafelski12} at higher $z$. We also include the metal-rich DLAs from \citet{Ma15}, \citet{Fynbo17}, \citet{Noterdaeme17}, and \citet{Noterdaeme10}. The latter is not strictly a DLA, but fulfils the $\log N({\rm \hi}) \geq 20$ criterion adopted in Paper I. The clean sample comprises 24 systems in total. The purpose of the clean sample is to monitor the corrections that we needed to apply to metallicities due to dust depletion, based on reliable relative abundances. However, this selection is not necessarily representative of the whole DLA population. In particular, the paucity of DLAs at high and low redshift makes it statistically incomplete. We therefore did not use this sample to assess the evolution of metallicity with redshift, but only to control the required dust corrections. To bypass this lack of completeness, we also considered the largest available DLA sample, below. Further potential biases are discussed in Sect. \ref{sec met evolution}.\\

\subsection{The large sample}
We applied the dust correction to as many DLA abundances as available in the literature, with sufficient relative abundances measurements to derive a sensible dust-correction. For this, we sourced the recent literature collection of \citet{Berg15}.\footnote{There is a small discrepancy between the $\log N({\rm \hi})$ for J1208+0010 between the values originally reported by \citet{Rafelski14} and later by \citet{Berg15}. We use the former value $\log N({\rm \hi})=20.30\pm0.15$, because regarded as more reliable (Berg, private communication). There are five potential duplicates inside the large sample: 1) J0035-0918; 2) QSO0201+36 and QSO0201+365; 3) J1340+1106 and Q1337+113, 4) B1036-2257 and Q1036-2257, and 5) J1356-1101 and Q1354-1046. These are likely due to different naming of the same quasars in different works. However, we did not attempt to investigate the origin of these potential duplicates and which measurements are the most reliable, but used the results as reported by \citet{Berg15}.}  We also included the metal-rich DLAs from \citet{Ma15}, \citet{Fynbo17}, \citet{Noterdaeme17}, \citet{Noterdaeme10}, and two $z\sim5$ absorbers from \citet{Poudel17}. Two of these additional systems are not strictly DLAs, but fulfil the $\log N({\rm \hi}) \geq 20$ criterion adopted in Paper I. The total number of DLAs in this sample is 236, which is comparable to the sample of \citet{Rafelski12} and its high-$z$ extension \citep{Rafelski14}. While the sample of \citet{Berg15} is originally much larger (almost 400 DLAs), we filtered those DLAs where Fe and at least another metal were well constrained to allow proper dust corrections. The purpose of the large sample is to provide dust-corrections for all literature DLAs, and furthermore to investigate the evolution of dust-corrected metallicity with redshift. Possible biases are discussed in Sect. \ref{sect bias}.

\section{The importance of dust corrections}
\label{sec importance dust corr}

      \begin{figure}
   \centering
   \includegraphics[width=9cm]{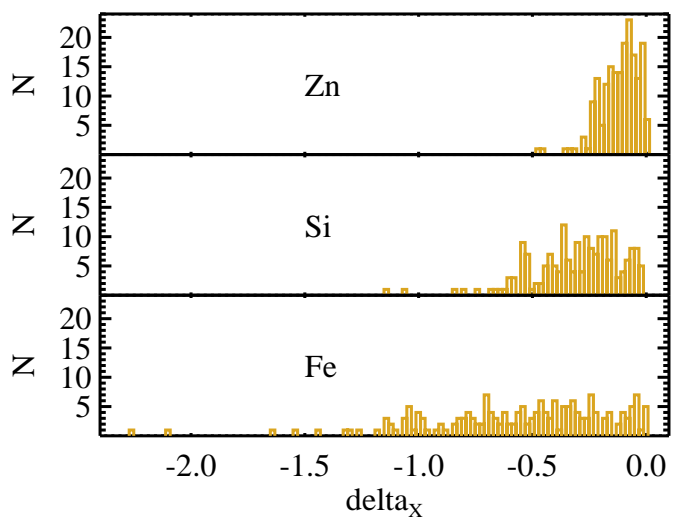}
   \caption{Distribution of the dust corrections (that is, depletions) for Zn, Si, and Fe that we calculate for the large sample.}
              \label{fig hist delta}
    \end{figure}
    \begin{figure*}[!h]
   \centering
   \includegraphics[width=18cm]{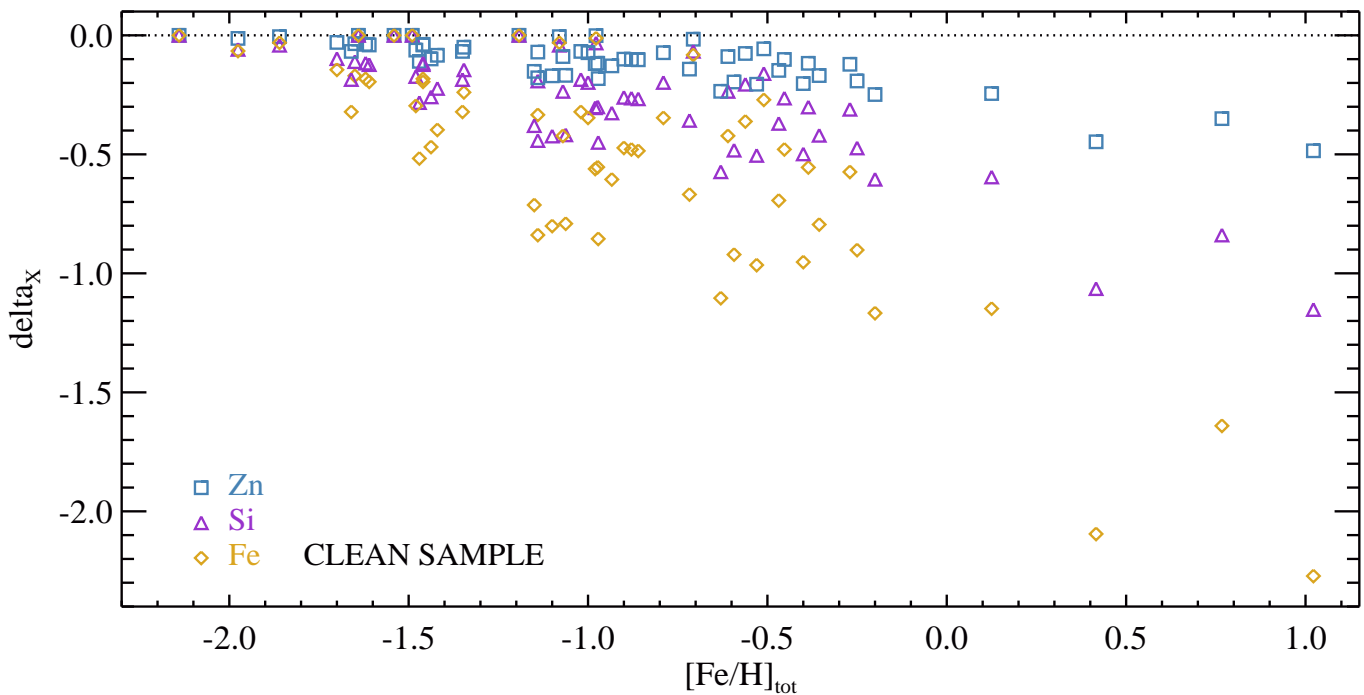}
   \caption{Depletions of Zn, Si, and Fe in the clean sample (Sect. \ref{sect sample}), with the dust-corrected metallicity [Fe/H]$_{\rm tot}$. Each individual system is represented by a value of $\delta_{\rm Zn}$, $\delta_{\rm Si}$, and $\delta_{\rm Fe}$.}
              \label{fig delta mh clean}
    \end{figure*}
Dust corrections are important, and crucial for metal-rich systems.  Figure \ref{fig hist delta} shows the overall distributions of depletions for Zn, Si, and Fe, in the large sample. Figure \ref{fig delta mh clean} shows the extent of the depletions of Zn, Si, and Fe in the clean sample with the dust-corrected metallicity. Each individual system is represented by a value of $\delta_{\rm Zn}$, $\delta_{\rm Si}$, and $\delta_{\rm Fe}$ in this plot. Clearly, the dust corrections increase with metallicity. Typically Zn is depleted by $\sim$0.1--0.2~dex in most DLAs, but depletes up to $0.5$~dex for the most metal-rich DLAs \citep{Ma15,Noterdaeme17}. As a comparison, in the Galaxy Zn depletes up to $0.67$~dex in the dustiest Galactic lines of sight \citep[$\zeta$ Oph, e.g.,] []{Savage96}. Therefore [Zn/H] is not a dust-free measurement of metallicity, and thus not necessarily a good metallicity estimator without dust corrections. Silicon and iron deplete up to $1.1$ and $2.30$~dex, respectively, in our samples. Even for less dusty, metal-poor end, Zn, Si, and even more Fe, can be still easily depleted, of about $\sim0.1$, $\sim0.3$, and $\sim0.5$~dex for systems with [Zn/Fe] $=0.4$, for Zn, Si, and Fe, respectively.While a significant fraction of DLAs have very little dust depletion, and therefore has a low dust and metal content \citep[as well as molecules, e.g.,] []{Petitjean00,Noterdaeme08,Ledoux09}, it is evident that systems with more dust and molecules do exist and are an important part of the DLA population. 

The [Fe/H]$_{\rm tot}$ is a measure of the total, dust-corrected metallicity, without influence of $\alpha$-element enhancement. These can be compared to stellar [Fe/H]. However, while stellar abundances typically have a wide range of abundances for a given galaxy \citep[e.g.,] []{Tolstoy09}, the ISM is the product of a longer-term reprocessing and recycling of metals in the gas, and has a narrower range of metallicities \citep[e.g.,] []{Krumholz17}. For the Galaxy, the neutral ISM metallicity is typically assumed to be solar \citep[e.g.,] []{Savage96,Jenkins09}. In addition, there may be some ISM metallicity gradients from the inner to the outer regions of the galaxies, which for DLAs should be shallow \citep[e.g., 0.022 dex/kpc][]{Christensen14}. The position of the knee of the $\alpha$-element enhancement distribution depends on the mass of the galaxy, where lower mass galaxies show the $\alpha$-element knee at lower metallicities \citet{Tolstoy09,deBoer14}. When dust depletion is taken into account, the intrinsic [$\alpha$/Fe] in DLAs are not strongly enhanced \citep[as also remarked by][]{Vladilo02}. $\alpha$-element enhancement levels similar to the Galaxy have been observed in DLAs with low-metallicity (Paper I). Further analysis on the [$\alpha$/Fe] in DLAs will be presented in De Cia et. al., in preparation.

\section{Metallicity evolution with cosmic time}
\label{sec met evolution}

\begin{figure*}[!h]
   \centering
   \includegraphics[width=18cm]{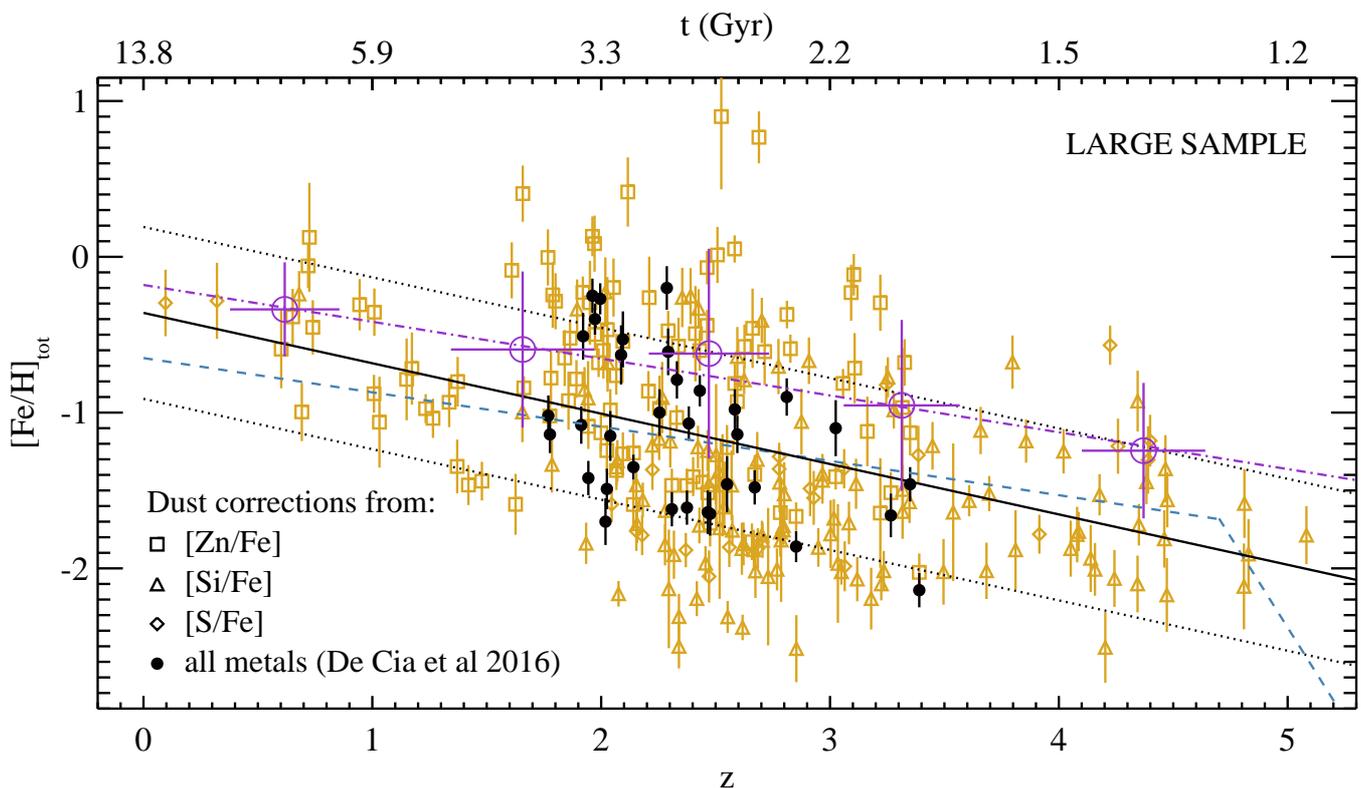}
   \caption{Dust-corrected metallicities [$M$/H]$_{\rm tot}$ derived from several metals simultaneously in Paper I (black filled circles), and the dust-corrected metallicities [Fe/H]$_{\rm tot}$ calculated with the single-reference method (Sect. \ref{sec dust corrections}) for the large sample (gold open symbols). The shape of the symbols show which reference was used for the dust correction, as labeled. The black solid and dotted lines display the linear fit to the data and the intrinsic scatter of the relation, respectively. The large open purple circles show the mean DLA metallicities weighted for the $N$(\hi) content, in bins of redshift, and the linear fit to these points is shown by the dotted-dashed purple line. The dashed blue curve shows the average DLA metallicity derived by \citet{Rafelski12} - originally derived for a scale of [Si/H] and weighted by the $N$(\hi) content - and the drop at high $z$ suggested by \citet{Rafelski14}}.
              \label{fig metz all}
              \end{figure*}

Figure \ref{fig metz all} shows the evolution of dust-corrected [Fe/H]$_{\rm tot}$ with $z$ for the DLAs of the large sample. The dust-corrected metallicities for the clean and the large sample are reported in Table \ref{tab met clean} and C.2, respectively. The Figure shows the solid metallicity measurements from Paper I, which were derived from several metal relative abundances simultaneously, and the [Fe/H]$_{\rm tot}$ that we derived as described in Sect. \ref{sec dust corrections}, where the dust corrections were calculated based on [Zn/Fe], or [Si/Fe], or [S/Fe], as labeled. We fit a linear relation to the data, where errors on both $x$ and $y$ data are considered, and including the intrinsic scatter $\sigma_{\rm int}$.\footnote{This is a linear least-squares approximation in one dimension ($y = a + b x$) that considers errors on $x$ and $y$ data ($\sigma_x$ and $\sigma_y$), using the \textsc{IDL} routine \textsc{MPFITEXY} \citep{Williams10}. \textsc{MPFITEXY} utilizes the \textsc{MPFIT} package \citep{Markwardt09}. Each data point is weighted as $1/\sqrt{\sigma_x^2 + b^2\sigma_y^2 + \sigma_{\rm int}^2}$, where $\sigma_{\rm int}$ is the intrinsic scatter of the data around the model \citep[“Nukers’ Estimate”;][]{Tremaine02}. The value $\sigma_{\rm int}$ is automatically scaled to produce a reduced $\chi^2_\nu\sim1$. We adopted an initial guess for $\sigma_{\rm int}$ of 0.1~dex.} The extent of the intrinsic scatter resulting from this fit is shown in Fig. \ref{fig metz all}. The error estimates of the linear fit to the data are correlated, because the majority of the data is centered around $z=2.5$. In this case, the uncertainty on the zero-intercept is $0.11$~dex, but suffers from this inter-dependence. A linear fit to the data, but with a displaced origin at $z=2.5$, does not suffer from this effect. In this case, the uncertainty on the zero-intercept is $0.04$~dex and we used this value. The results from the linear fit to the data are reported in Table \ref{tab fit}. 

We stress that the mean DLA metallicity is not necessarily the cosmic mean metallicity of the neutral gas, because the DLA sample may be incomplete. Biases and completeness are discussed in Sect. \ref{sect bias}. In addition, the cosmic mean metallicity of the neutral gas should be calculated by weighting the DLA metallicities for the $N$(\hi) content of each system, to avoid giving too much importance to the low-metallicity systems which carry less gas. Although our sample is unlikely complete at all redshifts, we calculated the mean DLA metallicity, weighted for the $N$(\hi) content, in bins of redshift ($z<1$, $1 \leq z < 2$, $2 \leq z < 3$, $ z \geq 4 $). The weighted metallicity is shown in Fig. \ref{fig metz all}, where the uncertainties are the standard deviations of the metallicities and redshifts of the DLAs in each redshift bin. We derived the linear fit to the weighted metallicities using the \textsc{MPFITEXY} fitting routine described above. The results from this fit are reported in Table \ref{tab fit}.

We compare our results with those of \citet{Rafelski12}, because it was the largest study of DLA metallicity evolution with $z$ until now. The mean DLA metallicity [Si/H] $= -0.65 + (-0.22 \times z)$ of \citet{Rafelski12} , shown in Fig. \ref{fig metz all}, was derived without dust-corrections and weighting the metallicities with the $N$(\hi) content. The drop at high redshift is an extrapolation of the results of \citet{Rafelski14}, which extended the analysis of \citet{Rafelski12} with a sample of high-$z$ DLAs. We stress that in Fig. \ref{fig metz all} we show a scale of [Fe/H]$_{\rm tot}$, while \citet{Rafelski12} and \citet{Rafelski14} used a scale of [Si/H]. These authors use either the observed Zn, Si or S metallicity, or otherwise the Fe metallicity enhanced by a systematic value of 0.3~dex to compensate for possible $\alpha$-element enhancement.\footnote{This also artificially compensates for some dust depletion by a mean depletion of iron of 0.3~dex into dust. However, this value is arbitrary and is far from the necessary dust corrections for Fe in DLAs, as we discuss below.} Thus, the dashed blue curve may in fact be lower than in Fig. \ref{fig metz all}. Nevertheless, we decided not to apply any shift, because the $\alpha$-element enhancement is likely metallicity dependent, and we did not make any assumption on [$\alpha$/Fe] in this work.   
 
We also compare our results with those of \citet{Vladilo02}, who studied the cosmic evolution of iron metallicity in a small sample of DLAs including independent dust corrections. We obtained the same slope of the metallicity evolution found by \citet{Vladilo02}, which is a reassuring result for both methods. The improvement with respect of the former work is more simplicity and less assumptions for the dust-correction method presented here, and the larger size of the DLA sample.  
 
The metallicity of DLAs decreases with $z$, and the correlation is significant, as confirmed by the correlation coefficients and low null-hypothesis probabilities (reported in Table \ref{tab fit}). The slope that we find is 0.1~dex per unit $z$ steeper than what was previously found by \citet{Rafelski12}, where no dust corrections were applied. \citet{Rafelski12} find a slope of $-0.22\pm 0.03$~dex (without dust corrections but weighting DLA metallicities by the \hi{} content) and we find $-0.32 \pm 0.04$~dex (with dust corrections but not weighting DLA metallicities by the \hi{} content). While the difference is not large, compared to the formal uncertainties in the slopes. Nevertheless, we note that a steeper slope can be expected with the inclusion of dust corrections, as also confirmed by \citet{Vladilo02} who found the same slope of $\sim -0.32$ with an independent method. A more consistent comparison with the work of \citep{Rafelski12} is found when weighting the DLA metallicities for the $N$(\hi) content of each system provides. In this case, we find a slope of the metallicity vs redshift relation of $-0.24\pm 0.14$~dex, which is very similar to \citet{Rafelski12}, but normalized to 0.4--0.5~dex higher metallicities than in that work, because of the dust corrections. 

In the analysis of our samples we find a significant difference with the previous results on the metallicity evolution of the neutral gas at low $z$, that is, about 0.3~dex higher metallicities than without taking dust corrections into considerations \citep{Rafelski12}, and up to 0.5~dex higher metallicities when weighting the mean DLA metallicity by the \hi{} content. This arises mostly from the fact that it is indeed in the metal-rich regime that dust depletion is strongest, and even for Zn and Si, which is often not considered. On the high-$z$ end, we do not find evidence for a steepening in the metallicity evolution, but more DLA measurements are needed to solidly compare to previous results \citep[e.g.,] []{Rafelski14}.

Interestingly, the evolution of metallicity with $z$ that can be found using DLAs (in absorption) is much steeper that what has been found for galaxies in emission, even taking different mass bins into account, while the low-$z$ results may be consistent \citep[e.g.,] []{Hunt16}. Determining the metallicity from emission lines may be challenging at high $z$ \citep[e.g.,] []{Kewley08}. However, we expect the gas probed by DLAs to be physically more extended than the classical ISM that is illuminated by stars in disks and shows the strong emission lines. Thus, the metallicities in absorption and emission do not have to necessarily agree. Comparing our results for the neutral gas with the metallicity evolution of the ionized gas, traced by lower \hi{} absorbers ($\log N(\mbox{\hi{}})<19$), we confirm that DLAs lack the very low metallicities observed in low \hi{} systems, which are thought to be related to the circumgalactic or intergalactic medium \citep{Lehner16}.

The relation of metallicity with redshift shows a large scatter, of about 0.5~dex, similarly to what had been previously found by, for example, \citet{Rafelski12} and \citet{Neeleman13}. At any given $z$, we expect that DLAs may select galaxies with a range of different masses and metallicities. \citet{Dvorkin15} calculated that environmental effects such as halo abundance, mass and stellar content produces a scatter in the relation of metallicity with redshift for DLAs of at least 0.25~dex in metallicity. Thus, the scatter of the metallicity vs redshift relation is physical, and it reflects a spread in metallicity \citep[and mass, due to the mass-metallicity relation, e.g.,] []{Tremonti04}. Galaxies with lower masses and metallicities are expected to lie in the lower envelope of the relation of metallicity with cosmic time, and galaxies with higher masses and metallicities in the upper envelope. Massive galaxies like the Milky Way are rare among DLAs, but they do exist \citep[e.g.,] []{Ma15,Noterdaeme17}. Interestingly, the upper envelope of the metallicity vs redshift relation at low $z$ reaches solar metallicities, and a few DLAs have supersolar metallicity at moderate $z$. For a discussion on the mass-metallicity relation in DLAs and its extensions see \citet{Ledoux06}, \citet{Prochaska08}, \citet{Moller13}, \citet{Neeleman13}, \citet{Christensen14}, and \citet{Arabsalmani15}. The extent of the scatter can in principle carry important information on these scaling relations. However, selection biases and incompleteness influence the extent of this scatter. For example, the fact that the scatter seems larger at $2\lesssim z \lesssim3.5$ is the effect of lower and higher $z$ ranges having less measurements due to an observational bias, as we discuss in the next Section.

In Fig. \ref{fig metz all} we show a linear fit to the metallicities with respect to redshift, and show the cosmic time scale, converted using the relation between time and redshift for a flat Universe \citep[e.g.,] []{Thomas00}, and assuming $H_0=67.8$ and $\Omega_M=0.308$ \citep{Planck16}. A linear fit to the data along the time axis does not provide a good description of the data, especially at low and high $z$, meaning that the distribution of metallicity with time is not linear\footnote{The $\chi^2$ of the linear fit of metallicity with redshift and time are comparable, because dominated by the large physical scatter of the relation. Nevertheless, the residuals of a linear fit of metallicity along the time axis show a trend and completely fail at reproducing low-$z$ metallicities.}. This is in accordance with cosmic chemical evolution models, which generally predict an evolution of metallicity with cosmic time that is linear with $z$, at least out to $z\sim4$ or so \citep[e.g.,] []{Pei95,Calura03,Tumlinson10,Matteucci12,Gioannini17}. In the linear time frame, this reflects a flattening of the increase of metallicity below $z\sim$ 1--2. Indeed, the cosmic starformation density (and, broadly speaking, the buildup of metals) increases with the age of the Universe and peaks around $z\sim$ 1--2 \citep[e.g.,] []{Madau98}. While we find overall agreement between our observations and some chemical evolution models \citep[e.g.,] []{Pei95,Tumlinson10}, a careful comparison with different models is beyond the scope of this paper and should be addressed in detail in the future.     
\begin{table*}[!h]
\centering
\caption{Coefficients of the linear fit ([Fe/H]$_{\rm tot} = A + B \times z$) to the data in Fig. \ref{fig metz all}. $\sigma_{\rm int}$ is the intrinsic scatter of the correlations. $r$ and $\rho$ are the Pearson and Spearman correlation coefficients, respectively, and are listed with their respective null-probability ($p_r$ and $p_\rho$).}
\begin{tabular}{ l | r r r r r r r r}
\hline \hline
\rule[-0.2cm]{0mm}{0.8cm}
 Sample & $A$ & $B$  & $\sigma_{\rm int}$ & $r$ & $p_r$ & $\rho$ & $p_{\rho}$  \\ 
\hline
               LARGE & $-0.36\pm 0.04$ & $-0.32\pm 0.04$ & $ 0.55$ & $-0.44$ & $4.E-14$ & $-0.47$ & $4.E-16 $ \\
LARGE, \hi{}-weighted & $-0.18\pm 0.21$ & $-0.24\pm 0.14$ & $ 0.10$ & $-0.98$ & $4.E-03$ & $-1.00$ & $0.E+00 $ \\
 \hline\hline
 \end{tabular}
\label{tab fit}
\end{table*}

\section{Biases, (in)completeness, and caveats}
\label{sect bias}

Through DLAs we probe the evolution of the metallicity of the neutral gas with cosmic time. The mean metallicities that we find are not necessarily representative of the mean metallicity of galaxies or the mean metallicity of the Universe, and they only refer to the neutral gas content of galaxies. In this section we discuss what are the possible biases that could affect our selection of DLAs with respect to the global population of DLAs.

Typical flux-limited studies of galaxies (that is, observing the stellar, gas, molecular, or dust emission) are biased toward the brightest objects. This is not true for DLAs, although potential dust obscuration may have an effect on the selection of the background QSOs themselves, as we discuss below. DLA-selected galaxies do not suffer from this bias against faintest targets, but instead from different and complementary biases. DLAs are selected from the cross-section of the gas around galaxies of different kinds, including low-mass faint galaxies. The gas in larger galaxies can extend to larger distances \citep[see][for how this impacts DLA selection and searches]{Fynbo08,Krogager17}, and thus impact parameters may be biased toward large values. This may in turn produce a bias in the observed metallicity, in case of strong radial gradients. However, such gradients should be shallow in DLAs \citep[e.g., 0.022 dex/kpc,][]{Christensen14}. In addition, low-mass galaxies are the most common in number \citep[e.g.,] []{Fontana04}.

As mentioned above, the selection of QSOs behind DLAs may be biased by the presence of dust in the absorbers. Indeed, if the background QSO is obscured by dust, it may simply be missed by the surveys. Some studies have suggested that this could a strong bias \citep[e.g.,] []{Vladilo05}. On the other hand, low reddening has been found in most systems, also when including dust-independent selection methods of the QSO \citep[radio or X-ray selections,][]{Ellison05,Vladilo08,Krogager16b}. Moderately highly reddened systems are rare but exist \citep{Noterdaeme09,Krogager16,Fynbo17}. The effect of dust obscuration acts more strongly on systems with higher dust columns, perhaps more dusty and metal-rich systems which are more common at low $z$. If we were missing some of these metal-rich and dusty systems at low-$z$, the metallicity evolution curve could be even steeper than what we find. However, at high $z$ the rest-frame UV emission of the background QSOs is more efficiently absorbed by dust than the low-$z$ rest-frame optical \citep[e.g.,] []{Savaglio15}. Thus, this bias may be stronger at high $z$, but on the other hand there is decreasing dust content at higher $z$, mitigating this effect. Overall, the dust obscuration bias should be small given what we have discussed above, but quantifying this effect is beyond the scope of this paper. \citet{Pontzen09} have quantified that only 7\% of DLAs may be missing due to dust obscuration bias.   

Another bias that may be dependent on $z$ is the metal selection. Because Zn is more difficult to measure due to its fairly weak lines, it is virtually never observed at high $z$ (never at $z>3.5$ in our large sample). On the high-$z$ end, the easiest and common metals to measure are Si and Fe. As we discussed in Sect. \ref{sec dust corrections}, there is some effect on the dust corrections calculated only based on the observed [Si/Fe], and these affect mainly the very low metallicity systems by underestimating their metallicity (making a slightly steeper metallicity evolution curve). In this paper we have corrected for this effect, but it is useful to keep this in mind. On the high-$z$ end, estimating \hi{} is increasingly challenging because of the Ly-$\alpha$ forest becoming more crowded. The \hi{} column density measurements must rely on the onset of the red dumping wing, and this may perhaps introduce large systematic uncertainties ($>0.3$~dex).

In addition, low-$z$ DLAs are selected in a different way than at high $z$. Indeed, Ly-$\alpha$ is redshifted into the optical range above $z\gtrsim 2$, and the DLA identification can be done from ground-based observations. On the other hand, UV observations are needed to identify low-$z$ DLAs, and these are often selected from \mgii{} absorbers \citep[e.g.,] []{Petitjean90,Rao00}, which may in principle have different overall properties than classical DLAs, such as higher metallicity. However, this effect seems negligible, that is, the \mgii{}-selected DLAs do not show higher metallicities than the non-\mgii{}-selected DLAs \citep{Rao17}. For \mgii{}-selected DLAs to represent the mean cosmic metallicity of the neutral gas, the distribution of their kinematics properties (or \mgii{} equivalent width) should resemble the true population \citep{Rao17}. In general, the derived mean metallicity can be considered the mean cosmic neutral-gas metallicity only for a complete sample, that is, if the selected DLA sample represents well the true population of DLAs (in terms of \mgii{} strength, kinematics, $N$(\hi{}) distribution, or mass). We attempt to calculate the mean cosmic metallicity of the neutral gas in galaxies by weighting the mean DLA metallicity by the \hi{} content of each system. This is only valid if the sample is complete.

\begin{figure}
   \centering
   \includegraphics[width=9cm]{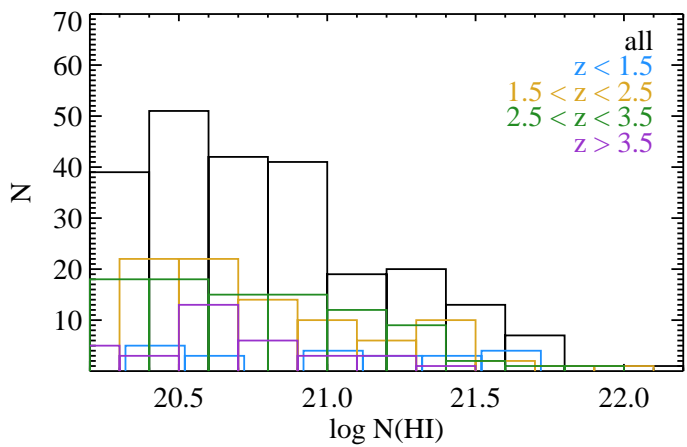}
   \caption{Distribution of $\log N$(\hi{}) for the large sample (all), and subsamples at particular redshift interval, as labeled.}
              \label{fig hist HI}
              \end{figure}

The completeness of the sample determines how well the given set of data represents the overall population of DLAs, and this also varies with $z$. Indeed, the majority of DLAs are observed at $2\lesssim z \lesssim3.5$. One way of assessing the completeness of our sample is by studying its $N$(\hi{}) distribution, which is shown in Fig. \ref{fig hist HI} for different redshift intervals. In the redshift range of about $1.5<z<3.5$ the $N$(\hi{}) distribution is similar to that of the large Sloan Digital Sky Survey (SDSS-DR5) DLA sample \citep{Prochaska05}, and of the Ultraviolet and Visual Echelle Spectrograph (UVES) DLA sample of \citet{Noterdaeme08}, which was scaled to compensate for completeness biases. In this $z$ interval we can therefore assume that the completeness is high, also given that other biases discussed above seem to be limited. On the other hand, at lower and higher $z$ the sample is still relatively small, and therefore we are missing the most extreme and rarest objects, on both the high- and low-metallicity envelopes of the metallicity evolution curve. For example high-$N$(\hi{}) systems are missing in at low and high $z$ (see Fig. \ref{fig hist HI}), although these systems are very rare \citep{Noterdaeme14}. Incompleteness is particularly severe at $z\lesssim0.5$ and $\gtrsim4.5$, where only a few measurements are available in our sample. The scatter of the metallicity evolution with redshift is heavily affected by incompleteness at high and low $z$, and we therefore refrain from studying it further in this paper. 

\section{Summary and conclusions}
\label{sect conclusions}
We developed a simplified method for calculating dust corrections to metal abundances (which we called the single-reference method), and confirmed the robustness of this method by comparing the dust corrections to the solid ones derived in Paper I by studying several metals simultaneously. We applied the new dust corrections to two DLA samples with published abundances: \textit{i)} a selection of high-quality measurements (the clean sample), and \textit{ii)} the largest number of available measurements (the large sample). From our analysis we conclude the following. 

\begin{enumerate}
 \item Dust corrections are important. Even Zn, which is often considered undepleted, can be depleted in DLAs, by typically $0.1$--$0.2$~dex and up to $0.5$~dex (Sect. \ref{sec importance dust corr}). 
 
\item Dust corrections are most crucial for more metal-rich systems. The depletions of Zn, Si, and Fe are shown in Fig. \ref{fig delta mh clean} as a function of dust-corrected metallicity.

 \item The DLA metallicities decrease with redshift. After including dust corrections, the slope of the metallicity decline with $z$ is steeper than what had previously found. Our best fit to the large sample yields [Fe/H]$_{\rm tot}=-0.36 - 0.32 \times z$, with a large internal scatter of 0.55~dex (Sect. \ref{sec met evolution}). When weighting the mean DLA metallicity by the \hi{} content, we find [Fe/H]$_{\rm tot}=-0.18 - 0.24 \times z$.

 \item The average DLA metallicity is 0.4--0.5~dex higher than what previously thought, when taking dust corrections into account.
 
 \item We do not find evidence for a steepening of the evolution of metallicity at high $z$. However, more measurements are needed to draw solid conclusions on the high-$z$ regime. 
  
 \item{We derived the cosmic evolution of dust-corrected metallicity of iron in the neutral gas, [Fe/H]$_{\rm tot}$. This scale carries no assumptions on the $\alpha$-element enhancement.}
 
 \item The scatter of the relation of metallicity with $z$ is physical. At any given $z$ a range of galaxies metallicities (and masses) is indeed expected. However, possible biases due to selection effects may affect the extent of the scatter. 
 
 \item The upper envelope of the relation of metallicity with $z$ reaches solar metallicity at low $z$.

\item We confirm that the DLA metallicity evolution with cosmic time supports the scenario where DLAs are associated with gas in and around galaxies with a wide range in metallicity and mass. While they have predominantly low metallicities and masses, DLAs can occasionally select also more metal-rich and massive systems.  

\item The dust-corrected metallicity of the neutral gas in galaxies decreases by a factor of $\sim50$--$100$ from today to $z=5$.

 \end{enumerate}

\begin{acknowledgements} 
We thank the referee, Prof. Edward Jenkins, for an extremely useful, insightful, and constructive report. We thank Pasquier Noterdaeme for insightful discussions and suggestions. This research has made use of NASA's Astrophysics Data System.
\end{acknowledgements}

 \bibliographystyle{aa} 
  
  \bibliography{biblio.bib}

\begin{thebibliography}{127}
\expandafter\ifx\csname natexlab\endcsname\relax\def\natexlab#1{#1}\fi

\bibitem[{{Arabsalmani} {et~al.}(2015){Arabsalmani}, {M{\o}ller}, {Fynbo},
  {Christensen}, {Freudling}, {Savaglio}, \& {Zafar}}]{Arabsalmani15}
{Arabsalmani}, M., {M{\o}ller}, P., {Fynbo}, J.~P.~U., {et~al.} 2015, \mnras,
  446, 990

\bibitem[{{Barbuy} {et~al.}(2015){Barbuy}, {Fria{\c c}a}, {da Silveira},
  {Hill}, {Zoccali}, {Minniti}, {Renzini}, {Ortolani}, \&
  {G{\'o}mez}}]{Barbuy15}
{Barbuy}, B., {Fria{\c c}a}, A.~C.~S., {da Silveira}, C.~R., {et~al.} 2015,
  \aap, 580, A40

\bibitem[{{Berg} {et~al.}(2015{\natexlab{a}}){Berg}, {Ellison}, {Prochaska},
  {Venn}, \& {Dessauges-Zavadsky}}]{Berg15}
{Berg}, T.~A.~M., {Ellison}, S.~L., {Prochaska}, J.~X., {Venn}, K.~A., \&
  {Dessauges-Zavadsky}, M. 2015{\natexlab{a}}, \mnras, 452, 4326

\bibitem[{{Berg} {et~al.}(2015{\natexlab{b}}){Berg}, {Neeleman}, {Prochaska},
  {Ellison}, \& {Wolfe}}]{Berg15b}
{Berg}, T.~A.~M., {Neeleman}, M., {Prochaska}, J.~X., {Ellison}, S.~L., \&
  {Wolfe}, A.~M. 2015{\natexlab{b}}, \pasp, 127, 167

\bibitem[{{Calura} {et~al.}(2003){Calura}, {Matteucci}, \&
  {Vladilo}}]{Calura03}
{Calura}, F., {Matteucci}, F., \& {Vladilo}, G. 2003, \mnras, 340, 59

\bibitem[{{Cardelli} {et~al.}(1993){Cardelli}, {Federman}, {Lambert}, \&
  {Theodosiou}}]{Cardelli93}
{Cardelli}, J.~A., {Federman}, S.~R., {Lambert}, D.~L., \& {Theodosiou}, C.~E.
  1993, \apjl, 416, L41

\bibitem[{{Christensen} {et~al.}(2014){Christensen}, {M{\o}ller}, {Fynbo}, \&
  {Zafar}}]{Christensen14}
{Christensen}, L., {M{\o}ller}, P., {Fynbo}, J.~P.~U., \& {Zafar}, T. 2014,
  \mnras, 445, 225

\bibitem[{{Cooke} {et~al.}(2010){Cooke}, {Pettini}, {Steidel}, {King}, {Rudie},
  \& {Rakic}}]{Cooke10}
{Cooke}, R., {Pettini}, M., {Steidel}, C.~C., {et~al.} 2010, \mnras, 409, 679

\bibitem[{{de Boer} {et~al.}(2014){de Boer}, {Belokurov}, {Beers}, \&
  {Lee}}]{deBoer14}
{de Boer}, T.~J.~L., {Belokurov}, V., {Beers}, T.~C., \& {Lee}, Y.~S. 2014,
  \mnras, 443, 658

\bibitem[{{De Cia} {et~al.}(2011){De Cia}, {Jakobsson}, {Bj{\"o}rnsson},
  {Vreeswijk}, {Dhillon}, {Marsh}, {Chapman}, {Fynbo}, {Ledoux}, {Littlefair},
  {Malesani}, {Schulze}, {Smette}, {Zafar}, \& {Gudmundsson}}]{DeCia11}
{De Cia}, A., {Jakobsson}, P., {Bj{\"o}rnsson}, G., {et~al.} 2011, \mnras, 412,
  2229

\bibitem[{{De Cia} {et~al.}(2012){De Cia}, {Ledoux}, {Fox}, {Vreeswijk},
  {Smette}, {Petitjean}, {Bj{\"o}rnsson}, {Fynbo}, {Hjorth}, \&
  {Jakobsson}}]{DeCia12}
{De Cia}, A., {Ledoux}, C., {Fox}, A.~J., {et~al.} 2012, \aap, 545, A64

\bibitem[{{De Cia} {et~al.}(2016){De Cia}, {Ledoux}, {Mattsson}, {Petitjean},
  {Srianand}, {Gavignaud}, \& {Jenkins}}]{DeCia16}
{De Cia}, A., {Ledoux}, C., {Mattsson}, L., {et~al.} 2016, \aap, 596, A97

\bibitem[{{De Cia} {et~al.}(2013){De Cia}, {Ledoux}, {Savaglio}, {Schady}, \&
  {Vreeswijk}}]{DeCia13}
{De Cia}, A., {Ledoux}, C., {Savaglio}, S., {Schady}, P., \& {Vreeswijk}, P.~M.
  2013, \aap, 560, A88

\bibitem[{{Dessauges-Zavadsky}(2008)}]{Dessauges-Zavadsky08}
{Dessauges-Zavadsky}, M. 2008, in IAU Symposium, Vol. 255, Low-Metallicity Star
  Formation: From the First Stars to Dwarf Galaxies, ed. L.~K. {Hunt}, S.~C.
  {Madden}, \& R.~{Schneider}, 121--128

\bibitem[{{Dessauges-Zavadsky} {et~al.}(2006){Dessauges-Zavadsky}, {Prochaska},
  {D'Odorico}, {Calura}, \& {Matteucci}}]{Dessauges-Zavadsky06}
{Dessauges-Zavadsky}, M., {Prochaska}, J.~X., {D'Odorico}, S., {Calura}, F., \&
  {Matteucci}, F. 2006, \aap, 445, 93

\bibitem[{{Draine}(2011)}]{Draine11b}
{Draine}, B.~T. 2011, {Physics of the Interstellar and Intergalactic Medium}
  (Princeton University Press)

\bibitem[{{Duffau} {et~al.}(2017){Duffau}, {Caffau}, {Sbordone}, {Bonifacio},
  {Andrievsky}, {Korotin}, {Babusiaux}, {Salvadori}, {Monaco}, {Francois},
  {Skuladottir}, {Bragaglia}, {Donati}, {Spina}, {Gallagher}, {Ludwig},
  {Christlieb}, {Hansen}, {Mott}, {Steffen}, {Zaggia}, {Blanco-Cuaresma},
  {Calura}, {Friel}, {Jimenez-Esteban}, {Koch}, {Magrini}, {Pancino}, {Tang},
  {Tautvaisiene}, {Vallenari}, {Hawkins}, {Gilmore}, {Randich}, {Feltzing},
  {Bensby}, {Flaccomio}, {Smiljanic}, {Bayo}, {Carraro}, {Casey}, {Costado},
  {Damiani}, {Franciosini}, {Hourihane}, {Jofre}, {Lardo}, {Lewis},
  {Morbidelli}, {Sousa}, \& {Worley}}]{Duffau17}
{Duffau}, S., {Caffau}, E., {Sbordone}, L., {et~al.} 2017, ArXiv e-prints
  [\eprint[arXiv]{1704.02981}]

\bibitem[{{Dutta} {et~al.}(2014){Dutta}, {Srianand}, {Rahmani}, {Petitjean},
  {Noterdaeme}, \& {Ledoux}}]{Dutta14}
{Dutta}, R., {Srianand}, R., {Rahmani}, H., {et~al.} 2014, \mnras, 440, 307

\bibitem[{{Dvorkin} {et~al.}(2015){Dvorkin}, {Silk}, {Vangioni}, {Petitjean},
  \& {Olive}}]{Dvorkin15}
{Dvorkin}, I., {Silk}, J., {Vangioni}, E., {Petitjean}, P., \& {Olive}, K.~A.
  2015, \mnras, 452, L36

\bibitem[{{Ellison} {et~al.}(2005){Ellison}, {Hall}, \& {Lira}}]{Ellison05}
{Ellison}, S.~L., {Hall}, P.~B., \& {Lira}, P. 2005, \aj, 130, 1345

\bibitem[{{Ellison} {et~al.}(2012){Ellison}, {Kanekar}, {Prochaska}, {Momjian},
  \& {Worseck}}]{Ellison12}
{Ellison}, S.~L., {Kanekar}, N., {Prochaska}, J.~X., {Momjian}, E., \&
  {Worseck}, G. 2012, \mnras, 424, 293

\bibitem[{{Field}(1974)}]{Field74}
{Field}, G.~B. 1974, \apj, 187, 453

\bibitem[{{Fontana} {et~al.}(2004){Fontana}, {Pozzetti}, {Donnarumma},
  {Renzini}, {Cimatti}, {Zamorani}, {Menci}, {Daddi}, {Giallongo}, {Mignoli},
  {Perna}, {Salimbeni}, {Saracco}, {Broadhurst}, {Cristiani}, {D'Odorico}, \&
  {Gilmozzi}}]{Fontana04}
{Fontana}, A., {Pozzetti}, L., {Donnarumma}, I., {et~al.} 2004, \aap, 424, 23

\bibitem[{{Fynbo} {et~al.}(2017){Fynbo}, {Krogager}, {Heintz}, {Geier},
  {M{\o}ller}, {Noterdaeme}, {Christensen}, {Ledoux}, \& {Jakobsson}}]{Fynbo17}
{Fynbo}, J.~P.~U., {Krogager}, J.-K., {Heintz}, K.~E., {et~al.} 2017, ArXiv
  e-prints [\eprint[arXiv]{1706.07016}]

\bibitem[{{Fynbo} {et~al.}(2008){Fynbo}, {Prochaska}, {Sommer-Larsen},
  {Dessauges-Zavadsky}, \& {M{\o}ller}}]{Fynbo08}
{Fynbo}, J.~P.~U., {Prochaska}, J.~X., {Sommer-Larsen}, J.,
  {Dessauges-Zavadsky}, M., \& {M{\o}ller}, P. 2008, \apj, 683, 321

\bibitem[{{Gioannini} {et~al.}(2017){Gioannini}, {Matteucci}, \&
  {Calura}}]{Gioannini17}
{Gioannini}, L., {Matteucci}, F., \& {Calura}, F. 2017, \mnras, 471, 4615

\bibitem[{{Henry} \& {Prochaska}(2007)}]{Henry07}
{Henry}, R.~B.~C. \& {Prochaska}, J.~X. 2007, \pasp, 119, 962

\bibitem[{{Hobbs} {et~al.}(1993){Hobbs}, {Welty}, {Morton}, {Spitzer}, \&
  {York}}]{Hobbs93}
{Hobbs}, L.~M., {Welty}, D.~E., {Morton}, D.~C., {Spitzer}, L., \& {York},
  D.~G. 1993, \apj, 411, 750

\bibitem[{{Hunt} {et~al.}(2016){Hunt}, {Dayal}, {Magrini}, \&
  {Ferrara}}]{Hunt16}
{Hunt}, L., {Dayal}, P., {Magrini}, L., \& {Ferrara}, A. 2016, \mnras, 463,
  2002

\bibitem[{{Jenkins}(1973)}]{Jenkins73}
{Jenkins}, E.~B. 1973, \apj, 181, 761

\bibitem[{{Jenkins}(2009)}]{Jenkins09}
{Jenkins}, E.~B. 2009, \apj, 700, 1299

\bibitem[{{Jenkins} {et~al.}(1986){Jenkins}, {Savage}, \&
  {Spitzer}}]{Jenkins86}
{Jenkins}, E.~B., {Savage}, B.~D., \& {Spitzer}, Jr., L. 1986, \apj, 301, 355

\bibitem[{{Jenkins} \& {Wallerstein}(2017)}]{Jenkins17}
{Jenkins}, E.~B. \& {Wallerstein}, G. 2017, \apj, 838, 85

\bibitem[{{Kewley} \& {Ellison}(2008)}]{Kewley08}
{Kewley}, L.~J. \& {Ellison}, S.~L. 2008, \apj, 681, 1183

\bibitem[{{Krogager} {et~al.}(2016{\natexlab{a}}){Krogager}, {Fynbo}, {Heintz},
  {Geier}, {Ledoux}, {M{\o}ller}, {Noterdaeme}, {Venemans}, \&
  {Vestergaard}}]{Krogager16b}
{Krogager}, J.-K., {Fynbo}, J.~P.~U., {Heintz}, K.~E., {et~al.}
  2016{\natexlab{a}}, \apj, 832, 49

\bibitem[{{Krogager} {et~al.}(2016{\natexlab{b}}){Krogager}, {Fynbo},
  {Noterdaeme}, {Zafar}, {M{\o}ller}, {Ledoux}, {Kr{\"u}hler}, \&
  {Stockton}}]{Krogager16}
{Krogager}, J.-K., {Fynbo}, J.~P.~U., {Noterdaeme}, P., {et~al.}
  2016{\natexlab{b}}, \mnras, 455, 2698

\bibitem[{{Krogager} {et~al.}(2017){Krogager}, {M{\o}ller}, {Fynbo}, \&
  {Noterdaeme}}]{Krogager17}
{Krogager}, J.-K., {M{\o}ller}, P., {Fynbo}, J.~P.~U., \& {Noterdaeme}, P.
  2017, \mnras, 469, 2959

\bibitem[{{Krumholz} \& {Ting}(2017)}]{Krumholz17}
{Krumholz}, M.~R. \& {Ting}, Y.-S. 2017, ArXiv e-prints
  [\eprint[arXiv]{1708.06853}]

\bibitem[{{Kulkarni} {et~al.}(1997){Kulkarni}, {Fall}, \&
  {Truran}}]{Kulkarni97}
{Kulkarni}, V.~P., {Fall}, S.~M., \& {Truran}, J.~W. 1997, \apjl, 484, L7

\bibitem[{{Kulkarni} {et~al.}(2007){Kulkarni}, {Khare}, {P{\'e}roux}, {York},
  {Lauroesch}, \& {Meiring}}]{Kulkarni07}
{Kulkarni}, V.~P., {Khare}, P., {P{\'e}roux}, C., {et~al.} 2007, \apj, 661, 88

\bibitem[{{Lanfranchi} \& {Fria{\c c}a}(2003)}]{Lanfranchi03}
{Lanfranchi}, G.~A. \& {Fria{\c c}a}, A.~C.~S. 2003, \mnras, 343, 481

\bibitem[{{Ledoux} {et~al.}(2002){Ledoux}, {Bergeron}, \&
  {Petitjean}}]{Ledoux02}
{Ledoux}, C., {Bergeron}, J., \& {Petitjean}, P. 2002, \aap, 385, 802

\bibitem[{{Ledoux} {et~al.}(2003){Ledoux}, {Petitjean}, \&
  {Srianand}}]{Ledoux03}
{Ledoux}, C., {Petitjean}, P., \& {Srianand}, R. 2003, \mnras, 346, 209

\bibitem[{{Ledoux} {et~al.}(2006){Ledoux}, {Petitjean}, \&
  {Srianand}}]{Ledoux06}
{Ledoux}, C., {Petitjean}, P., \& {Srianand}, R. 2006, \apjl, 640, L25

\bibitem[{{Ledoux} {et~al.}(2009){Ledoux}, {Vreeswijk}, {Smette}, {Fox},
  {Petitjean}, {Ellison}, {Fynbo}, \& {Savaglio}}]{Ledoux09}
{Ledoux}, C., {Vreeswijk}, P.~M., {Smette}, A., {et~al.} 2009, \aap, 506, 661

\bibitem[{{Lehner} {et~al.}(2015){Lehner}, {Howk}, \& {Wakker}}]{Lehner15}
{Lehner}, N., {Howk}, J.~C., \& {Wakker}, B.~P. 2015, \apj, 804, 79

\bibitem[{{Lehner} {et~al.}(2016){Lehner}, {O'Meara}, {Howk}, {Prochaska}, \&
  {Fumagalli}}]{Lehner16}
{Lehner}, N., {O'Meara}, J.~M., {Howk}, J.~C., {Prochaska}, J.~X., \&
  {Fumagalli}, M. 2016, \apj, 833, 283

\bibitem[{{Lelli} {et~al.}(2016){Lelli}, {McGaugh}, \& {Schombert}}]{Lelli16}
{Lelli}, F., {McGaugh}, S.~S., \& {Schombert}, J.~M. 2016, \aj, 152, 157

\bibitem[{{Ma} {et~al.}(2015){Ma}, {Caucal}, {Noterdaeme}, {Ge}, {Prochaska},
  {Ji}, {Zhang}, {Rahmani}, {Jiang}, {Schneider}, {Lundgren}, \&
  {P{\^a}ris}}]{Ma15}
{Ma}, J., {Caucal}, P., {Noterdaeme}, P., {et~al.} 2015, \mnras, 454, 1751

\bibitem[{{Madau} {et~al.}(1998){Madau}, {Pozzetti}, \& {Dickinson}}]{Madau98}
{Madau}, P., {Pozzetti}, L., \& {Dickinson}, M. 1998, \apj, 498, 106

\bibitem[{{Marasco} \& {Fraternali}(2011)}]{Marasco11}
{Marasco}, A. \& {Fraternali}, F. 2011, \aap, 525, A134

\bibitem[{{Markwardt}(2009)}]{Markwardt09}
{Markwardt}, C.~B. 2009, in Astronomical Society of the Pacific Conference
  Series, Vol. 411, Astronomical Data Analysis Software and Systems XVIII, ed.
  D.~A. {Bohlender}, D.~{Durand}, \& P.~{Dowler}, 251

\bibitem[{{Matteucci}(2012)}]{Matteucci12}
{Matteucci}, F. 2012, {Chemical Evolution of Galaxies}

\bibitem[{{Meiring} {et~al.}(2006){Meiring}, {Kulkarni}, {Khare}, {Bechtold},
  {York}, {Cui}, {Lauroesch}, {Crotts}, \& {Nakamura}}]{Meiring06}
{Meiring}, J.~D., {Kulkarni}, V.~P., {Khare}, P., {et~al.} 2006, \mnras, 370,
  43

\bibitem[{{Meiring} {et~al.}(2011){Meiring}, {Tripp}, {Prochaska}, {Tumlinson},
  {Werk}, {Jenkins}, {Thom}, {O'Meara}, \& {Sembach}}]{Meiring11}
{Meiring}, J.~D., {Tripp}, T.~M., {Prochaska}, J.~X., {et~al.} 2011, \apj, 732,
  35

\bibitem[{{Milutinovic} {et~al.}(2010){Milutinovic}, {Ellison}, {Prochaska}, \&
  {Tumlinson}}]{Milutinovic10}
{Milutinovic}, N., {Ellison}, S.~L., {Prochaska}, J.~X., \& {Tumlinson}, J.
  2010, \mnras, 408, 2071

\bibitem[{{M{\o}ller} {et~al.}(2013){M{\o}ller}, {Fynbo}, {Ledoux}, \&
  {Nilsson}}]{Moller13}
{M{\o}ller}, P., {Fynbo}, J.~P.~U., {Ledoux}, C., \& {Nilsson}, K.~K. 2013,
  \mnras, 430, 2680

\bibitem[{{Morton}(1975)}]{Morton75}
{Morton}, D.~C. 1975, \apj, 197, 85

\bibitem[{{Neeleman} {et~al.}(2013){Neeleman}, {Wolfe}, {Prochaska}, \&
  {Rafelski}}]{Neeleman13}
{Neeleman}, M., {Wolfe}, A.~M., {Prochaska}, J.~X., \& {Rafelski}, M. 2013,
  \apj, 769, 54

\bibitem[{{Nissen} {et~al.}(2007){Nissen}, {Akerman}, {Asplund}, {Fabbian},
  {Kerber}, {Kaufl}, \& {Pettini}}]{Nissen07}
{Nissen}, P.~E., {Akerman}, C., {Asplund}, M., {et~al.} 2007, \aap, 469, 319

\bibitem[{{Nissen} \& {Schuster}(2011)}]{Nissen11}
{Nissen}, P.~E. \& {Schuster}, W.~J. 2011, \aap, 530, A15

\bibitem[{{Nomoto} {et~al.}(1997){Nomoto}, {Iwamoto}, {Nakasato}, {Thielemann},
  {Brachwitz}, {Tsujimoto}, {Kubo}, \& {Kishimoto}}]{Nomoto97}
{Nomoto}, K., {Iwamoto}, K., {Nakasato}, N., {et~al.} 1997, Nuclear Physics A,
  621, 467

\bibitem[{{Noterdaeme} {et~al.}(2017){Noterdaeme}, {Krogager}, {Balashev},
  {Ge}, {Gupta}, {Kr{\"u}hler}, {Ledoux}, {Murphy}, {P{\^a}ris}, {Petitjean},
  {Rahmani}, {Srianand}, \& {Ubachs}}]{Noterdaeme17}
{Noterdaeme}, P., {Krogager}, J.-K., {Balashev}, S., {et~al.} 2017, \aap, 597,
  A82

\bibitem[{{Noterdaeme} {et~al.}(2008){Noterdaeme}, {Ledoux}, {Petitjean}, \&
  {Srianand}}]{Noterdaeme08}
{Noterdaeme}, P., {Ledoux}, C., {Petitjean}, P., \& {Srianand}, R. 2008, \aap,
  481, 327

\bibitem[{{Noterdaeme} {et~al.}(2009){Noterdaeme}, {Ledoux}, {Srianand},
  {Petitjean}, \& {Lopez}}]{Noterdaeme09}
{Noterdaeme}, P., {Ledoux}, C., {Srianand}, R., {Petitjean}, P., \& {Lopez}, S.
  2009, \aap, 503, 765

\bibitem[{{Noterdaeme} {et~al.}(2010){Noterdaeme}, {Petitjean}, {Ledoux},
  {L{\'o}pez}, {Srianand}, \& {Vergani}}]{Noterdaeme10}
{Noterdaeme}, P., {Petitjean}, P., {Ledoux}, C., {et~al.} 2010, \aap, 523, A80

\bibitem[{{Noterdaeme} {et~al.}(2014){Noterdaeme}, {Petitjean}, {P{\^a}ris},
  {Cai}, {Finley}, {Ge}, {Pieri}, \& {York}}]{Noterdaeme14}
{Noterdaeme}, P., {Petitjean}, P., {P{\^a}ris}, I., {et~al.} 2014, \aap, 566,
  A24

\bibitem[{{O'Meara} {et~al.}(2006){O'Meara}, {Burles}, {Prochaska}, {Prochter},
  {Bernstein}, \& {Burgess}}]{OMeara06}
{O'Meara}, J.~M., {Burles}, S., {Prochaska}, J.~X., {et~al.} 2006, \apjl, 649,
  L61

\bibitem[{{Pei} \& {Fall}(1995)}]{Pei95}
{Pei}, Y.~C. \& {Fall}, S.~M. 1995, \apj, 454, 69

\bibitem[{{P{\'e}roux} {et~al.}(2007){P{\'e}roux}, {Dessauges-Zavadsky},
  {D'Odorico}, {Kim}, \& {McMahon}}]{Peroux07}
{P{\'e}roux}, C., {Dessauges-Zavadsky}, M., {D'Odorico}, S., {Kim}, T.-S., \&
  {McMahon}, R.~G. 2007, \mnras, 382, 177

\bibitem[{{P{\'e}roux} {et~al.}(2006){P{\'e}roux}, {Meiring}, {Kulkarni},
  {Ferlet}, {Khare}, {Lauroesch}, {Vladilo}, \& {York}}]{Peroux06}
{P{\'e}roux}, C., {Meiring}, J.~D., {Kulkarni}, V.~P., {et~al.} 2006, \mnras,
  372, 369

\bibitem[{{P{\'e}roux} {et~al.}(2008){P{\'e}roux}, {Meiring}, {Kulkarni},
  {Khare}, {Lauroesch}, {Vladilo}, \& {York}}]{Peroux08}
{P{\'e}roux}, C., {Meiring}, J.~D., {Kulkarni}, V.~P., {et~al.} 2008, \mnras,
  386, 2209

\bibitem[{{Petitjean} \& {Bergeron}(1990)}]{Petitjean90}
{Petitjean}, P. \& {Bergeron}, J. 1990, \aap, 231, 309

\bibitem[{{Petitjean} {et~al.}(2000){Petitjean}, {Srianand}, \&
  {Ledoux}}]{Petitjean00}
{Petitjean}, P., {Srianand}, R., \& {Ledoux}, C. 2000, \aap, 364, L26

\bibitem[{{Pettini} {et~al.}(1999){Pettini}, {Ellison}, {Steidel}, \&
  {Bowen}}]{Pettini99}
{Pettini}, M., {Ellison}, S.~L., {Steidel}, C.~C., \& {Bowen}, D.~V. 1999,
  \apj, 510, 576

\bibitem[{{Pettini} {et~al.}(2000){Pettini}, {Ellison}, {Steidel}, {Shapley},
  \& {Bowen}}]{Pettini00}
{Pettini}, M., {Ellison}, S.~L., {Steidel}, C.~C., {Shapley}, A.~E., \&
  {Bowen}, D.~V. 2000, \apj, 532, 65

\bibitem[{{Pettini} {et~al.}(1994){Pettini}, {Smith}, {Hunstead}, \&
  {King}}]{Pettini94}
{Pettini}, M., {Smith}, L.~J., {Hunstead}, R.~W., \& {King}, D.~L. 1994, \apj,
  426, 79

\bibitem[{{Phillips} {et~al.}(1982){Phillips}, {Gondhalekar}, \&
  {Pettini}}]{Phillips82}
{Phillips}, A.~P., {Gondhalekar}, P.~M., \& {Pettini}, M. 1982, \mnras, 200,
  687

\bibitem[{{Phillips} {et~al.}(1984){Phillips}, {Pettini}, \&
  {Gondhalekar}}]{Phillips84}
{Phillips}, A.~P., {Pettini}, M., \& {Gondhalekar}, P.~M. 1984, \mnras, 206,
  337

\bibitem[{{Planck Collaboration} {et~al.}(2016){Planck Collaboration}, {Ade},
  {Aghanim}, {Arnaud}, {Ashdown}, {Aumont}, {Baccigalupi}, {Banday},
  {Barreiro}, {Bartlett}, \& et~al.}]{Planck16}
{Planck Collaboration}, {Ade}, P.~A.~R., {Aghanim}, N., {et~al.} 2016, \aap,
  594, A13

\bibitem[{{Pontzen} {et~al.}(2008){Pontzen}, {Governato}, {Pettini}, {Booth},
  {Stinson}, {Wadsley}, {Brooks}, {Quinn}, \& {Haehnelt}}]{Pontzen08}
{Pontzen}, A., {Governato}, F., {Pettini}, M., {et~al.} 2008, \mnras, 390, 1349

\bibitem[{{Pontzen} \& {Pettini}(2009)}]{Pontzen09}
{Pontzen}, A. \& {Pettini}, M. 2009, \mnras, 393, 557

\bibitem[{{Poudel} {et~al.}(2017){Poudel}, {Kulkarni}, {Morrison},
  {P{\'e}roux}, {Som}, {Rahmani}, \& {Quiret}}]{Poudel17}
{Poudel}, S., {Kulkarni}, V.~P., {Morrison}, S., {et~al.} 2017, ArXiv e-prints
  [\eprint[arXiv]{1710.03315}]

\bibitem[{{Primas} {et~al.}(2000){Primas}, {Brugamyer}, {Sneden}, {King},
  {Beers}, {Boesgaard}, \& {Deliyannis}}]{Primas00}
{Primas}, F., {Brugamyer}, E., {Sneden}, C., {et~al.} 2000, in The First Stars,
  ed. A.~{Weiss}, T.~G. {Abel}, \& V.~{Hill}, 51

\bibitem[{{Prochaska} {et~al.}(2008){Prochaska}, {Chen}, {Wolfe},
  {Dessauges-Zavadsky}, \& {Bloom}}]{Prochaska08}
{Prochaska}, J.~X., {Chen}, H., {Wolfe}, A.~M., {Dessauges-Zavadsky}, M., \&
  {Bloom}, J.~S. 2008, \apj, 672, 59

\bibitem[{{Prochaska} {et~al.}(2003){Prochaska}, {Gawiser}, {Wolfe}, {Castro},
  \& {Djorgovski}}]{Prochaska03}
{Prochaska}, J.~X., {Gawiser}, E., {Wolfe}, A.~M., {Castro}, S., \&
  {Djorgovski}, S.~G. 2003, \apjl, 595, L9

\bibitem[{{Prochaska} {et~al.}(2002){Prochaska}, {Henry}, {O'Meara}, {Tytler},
  {Wolfe}, {Kirkman}, {Lubin}, \& {Suzuki}}]{Prochaska02b}
{Prochaska}, J.~X., {Henry}, R.~B.~C., {O'Meara}, J.~M., {et~al.} 2002, \pasp,
  114, 933

\bibitem[{{Prochaska} {et~al.}(2005){Prochaska}, {Herbert-Fort}, \&
  {Wolfe}}]{Prochaska05}
{Prochaska}, J.~X., {Herbert-Fort}, S., \& {Wolfe}, A.~M. 2005, \apj, 635, 123

\bibitem[{{Prochaska} \& {Wolfe}(2002)}]{Prochaska02}
{Prochaska}, J.~X. \& {Wolfe}, A.~M. 2002, \apj, 566, 68

\bibitem[{{Prochaska} {et~al.}(2001){Prochaska}, {Wolfe}, {Tytler}, {Burles},
  {Cooke}, {Gawiser}, {Kirkman}, {O'Meara}, \&
  {Storrie-Lombardi}}]{Prochaska01}
{Prochaska}, J.~X., {Wolfe}, A.~M., {Tytler}, D., {et~al.} 2001, \apjs, 137, 21

\bibitem[{{Quiret} {et~al.}(2016){Quiret}, {P{\'e}roux}, {Zafar}, {Kulkarni},
  {Jenkins}, {Milliard}, {Rahmani}, {Popping}, {Rao}, {Turnshek}, \&
  {Monier}}]{Quiret16}
{Quiret}, S., {P{\'e}roux}, C., {Zafar}, T., {et~al.} 2016, \mnras, 458, 4074

\bibitem[{{Rafelski} {et~al.}(2014){Rafelski}, {Neeleman}, {Fumagalli},
  {Wolfe}, \& {Prochaska}}]{Rafelski14}
{Rafelski}, M., {Neeleman}, M., {Fumagalli}, M., {Wolfe}, A.~M., \&
  {Prochaska}, J.~X. 2014, \apjl, 782, L29

\bibitem[{{Rafelski} {et~al.}(2012){Rafelski}, {Wolfe}, {Prochaska},
  {Neeleman}, \& {Mendez}}]{Rafelski12}
{Rafelski}, M., {Wolfe}, A.~M., {Prochaska}, J.~X., {Neeleman}, M., \&
  {Mendez}, A.~J. 2012, \apj, 755, 89

\bibitem[{{Rao} {et~al.}(2005){Rao}, {Prochaska}, {Howk}, \& {Wolfe}}]{Rao05}
{Rao}, S.~M., {Prochaska}, J.~X., {Howk}, J.~C., \& {Wolfe}, A.~M. 2005, \aj,
  129, 9

\bibitem[{{Rao} \& {Turnshek}(2000)}]{Rao00}
{Rao}, S.~M. \& {Turnshek}, D.~A. 2000, \apjs, 130, 1

\bibitem[{{Rao} {et~al.}(2017){Rao}, {Turnshek}, {Sardane}, \&
  {Monier}}]{Rao17}
{Rao}, S.~M., {Turnshek}, D.~A., {Sardane}, G.~M., \& {Monier}, E.~M. 2017,
  \mnras, 471, 3428

\bibitem[{{Rodr{\'{\i}}guez} {et~al.}(2006){Rodr{\'{\i}}guez}, {Petitjean},
  {Aracil}, {Ledoux}, \& {Srianand}}]{Rodriguez06}
{Rodr{\'{\i}}guez}, E., {Petitjean}, P., {Aracil}, B., {Ledoux}, C., \&
  {Srianand}, R. 2006, \aap, 446, 791

\bibitem[{{Saito} {et~al.}(2009){Saito}, {Takada-Hidai}, {Honda}, \&
  {Takeda}}]{Saito09}
{Saito}, Y.-J., {Takada-Hidai}, M., {Honda}, S., \& {Takeda}, Y. 2009, \pasj,
  61, 549

\bibitem[{{Savage} {et~al.}(1992){Savage}, {Cardelli}, \& {Sofia}}]{Savage92}
{Savage}, B.~D., {Cardelli}, J.~A., \& {Sofia}, U.~J. 1992, \apj, 401, 706

\bibitem[{{Savage} \& {Sembach}(1991)}]{Savage91}
{Savage}, B.~D. \& {Sembach}, K.~R. 1991, \apj, 379, 245

\bibitem[{{Savage} \& {Sembach}(1996)}]{Savage96}
{Savage}, B.~D. \& {Sembach}, K.~R. 1996, \araa, 34, 279

\bibitem[{{Savaglio}(2001)}]{Savaglio01}
{Savaglio}, S. 2001, in IAU Symposium, Vol. 204, The Extragalactic Infrared
  Background and its Cosmological Implications, ed. M.~{Harwit} \& M.~G.
  {Hauser}, 307

\bibitem[{{Savaglio}(2006)}]{Savaglio06}
{Savaglio}, S. 2006, New Journal of Physics, 8, 195

\bibitem[{{Savaglio}(2015)}]{Savaglio15}
{Savaglio}, S. 2015, Journal of High Energy Astrophysics, 7, 95

\bibitem[{{Sk{\'u}lad{\'o}ttir} {et~al.}(2017){Sk{\'u}lad{\'o}ttir}, {Tolstoy},
  {Salvadori}, {Hill}, \& {Pettini}}]{Skuladottir17}
{Sk{\'u}lad{\'o}ttir}, {\'A}., {Tolstoy}, E., {Salvadori}, S., {Hill}, V., \&
  {Pettini}, M. 2017, ArXiv e-prints [\eprint[arXiv]{1708.00511}]

\bibitem[{{Sneden} {et~al.}(1991){Sneden}, {Gratton}, \& {Crocker}}]{Sneden91}
{Sneden}, C., {Gratton}, R.~G., \& {Crocker}, D.~A. 1991, \aap, 246, 354

\bibitem[{{Som} {et~al.}(2013){Som}, {Kulkarni}, {Meiring}, {York},
  {P{\'e}roux}, {Khare}, \& {Lauroesch}}]{Som13}
{Som}, D., {Kulkarni}, V.~P., {Meiring}, J., {et~al.} 2013, \mnras, 435, 1469

\bibitem[{{Thomas} \& {Kantowski}(2000)}]{Thomas00}
{Thomas}, R.~C. \& {Kantowski}, R. 2000, \prd, 62, 103507

\bibitem[{{Tolstoy} {et~al.}(2009){Tolstoy}, {Hill}, \& {Tosi}}]{Tolstoy09}
{Tolstoy}, E., {Hill}, V., \& {Tosi}, M. 2009, \araa, 47, 371

\bibitem[{{Tremaine} {et~al.}(2002){Tremaine}, {Gebhardt}, {Bender}, {Bower},
  {Dressler}, {Faber}, {Filippenko}, {Green}, {Grillmair}, {Ho}, {Kormendy},
  {Lauer}, {Magorrian}, {Pinkney}, \& {Richstone}}]{Tremaine02}
{Tremaine}, S., {Gebhardt}, K., {Bender}, R., {et~al.} 2002, \apj, 574, 740

\bibitem[{{Tremonti} {et~al.}(2004){Tremonti}, {Heckman}, {Kauffmann},
  {Brinchmann}, {Charlot}, {White}, {Seibert}, {Peng}, {Schlegel}, {Uomoto},
  {Fukugita}, \& {Brinkmann}}]{Tremonti04}
{Tremonti}, C.~A., {Heckman}, T.~M., {Kauffmann}, G., {et~al.} 2004, \apj, 613,
  898

\bibitem[{{Tumlinson}(2010)}]{Tumlinson10}
{Tumlinson}, J. 2010, \apj, 708, 1398

\bibitem[{{Tumlinson} {et~al.}(2017){Tumlinson}, {Peeples}, \&
  {Werk}}]{Tumlinson17}
{Tumlinson}, J., {Peeples}, M.~S., \& {Werk}, J.~K. 2017, \araa, 55, 389

\bibitem[{{Viegas}(1995)}]{Viegas95}
{Viegas}, S.~M. 1995, \mnras, 276, 268

\bibitem[{{Vladilo}(1998)}]{Vladilo98}
{Vladilo}, G. 1998, \apj, 493, 583

\bibitem[{{Vladilo}(2002)}]{Vladilo02}
{Vladilo}, G. 2002, \aap, 391, 407

\bibitem[{{Vladilo} {et~al.}(2011){Vladilo}, {Abate}, {Yin}, {Cescutti}, \&
  {Matteucci}}]{Vladilo11}
{Vladilo}, G., {Abate}, C., {Yin}, J., {Cescutti}, G., \& {Matteucci}, F. 2011,
  \aap, 530, A33

\bibitem[{{Vladilo} {et~al.}(2001){Vladilo}, {Centuri{\'o}n}, {Bonifacio}, \&
  {Howk}}]{Vladilo01}
{Vladilo}, G., {Centuri{\'o}n}, M., {Bonifacio}, P., \& {Howk}, J.~C. 2001,
  \apj, 557, 1007

\bibitem[{{Vladilo} \& {P{\'e}roux}(2005)}]{Vladilo05}
{Vladilo}, G. \& {P{\'e}roux}, C. 2005, \aap, 444, 461

\bibitem[{{Vladilo} {et~al.}(2008){Vladilo}, {Prochaska}, \&
  {Wolfe}}]{Vladilo08}
{Vladilo}, G., {Prochaska}, J.~X., \& {Wolfe}, A.~M. 2008, \aap, 478, 701

\bibitem[{{Vreeswijk} {et~al.}(2013){Vreeswijk}, {Ledoux}, {Raassen}, {Smette},
  {De Cia}, {Wo{\'z}niak}, {Fox}, {Vestrand}, \& {Jakobsson}}]{Vreeswijk13}
{Vreeswijk}, P.~M., {Ledoux}, C., {Raassen}, A.~J.~J., {et~al.} 2013, \aap,
  549, A22

\bibitem[{{Welty} {et~al.}(1995){Welty}, {Hobbs}, {Lauroesch}, {Morton}, \&
  {York}}]{Welty95}
{Welty}, D.~E., {Hobbs}, L.~M., {Lauroesch}, J.~T., {Morton}, D.~C., \& {York},
  D.~G. 1995, \apjl, 449, L135

\bibitem[{{Williams} {et~al.}(2010){Williams}, {Bureau}, \&
  {Cappellari}}]{Williams10}
{Williams}, M.~J., {Bureau}, M., \& {Cappellari}, M. 2010, \mnras, 409, 1330

\bibitem[{{Wiseman} {et~al.}(2017){Wiseman}, {Schady}, {Bolmer}, {Kr{\"u}hler},
  {Yates}, {Greiner}, \& {Fynbo}}]{Wiseman17}
{Wiseman}, P., {Schady}, P., {Bolmer}, J., {et~al.} 2017, \aap, 599, A24

\bibitem[{{Wolfe} {et~al.}(2005){Wolfe}, {Gawiser}, \& {Prochaska}}]{Wolfe05}
{Wolfe}, A.~M., {Gawiser}, E., \& {Prochaska}, J.~X. 2005, \araa, 43, 861

\bibitem[{{Wolfe} {et~al.}(1995){Wolfe}, {Lanzetta}, {Foltz}, \&
  {Chaffee}}]{Wolfe95}
{Wolfe}, A.~M., {Lanzetta}, K.~M., {Foltz}, C.~B., \& {Chaffee}, F.~H. 1995,
  \apj, 454, 698

\bibitem[{{Zafar} {et~al.}(2014){Zafar}, {Centuri{\'o}n}, {P{\'e}roux},
  {Molaro}, {D'Odorico}, {Vladilo}, \& {Popping}}]{Zafar14b}
{Zafar}, T., {Centuri{\'o}n}, M., {P{\'e}roux}, C., {et~al.} 2014, \mnras, 444,
  744

\end{thebibliography}

\onecolumn
\appendix

\section{Potential need for an update of the depletion-sequences coefficients for Si}

One possible reason for the observed deviation of Si-based single-reference metallicities from the real metallicity (Fig. \ref{fig test met}) is a potential inaccurate estimate of the Si coefficients $A1_{\rm Si}$ and $B1_{\rm Si}$ in Paper I. This is not unlikely, given that the fit of the [Si/Zn] vs [Zn/Fe] relation was constrained by only few data points for the Galaxy, in fact only two datapoints at [Zn/Fe] $>1$, see top panels in Fig. 3 of Paper I. A steeper relation [Si/Zn] vs [Zn/Fe] could compensate for the observed trend of Si-based single-reference metallicities in Fig. \ref{fig test met}, namely for $A1_{\rm Si, new}=+0.43$ and $B1_{\rm Si, new}=-0.84$. Furthermore, this would imply that an $\alpha$-elements zero point for Si $\alpha_{Si, 0}=0.43$~dex, and a slope of the depletion sequence for Si $B2_{\rm Si, new}=-0.97$. However, more data at high [Zn/Fe] are necessary to assess whether the Si depletion sequences should be updated to these values. Other factors may play a role in producing the trend of Si-based single-reference metallicities (Fig. \ref{fig test met}).

\section{Inspection of literature values for DLAs with negative [Zn/Fe]$_{\rm exp}$}

We closely inspected all the DLAs in the large sample with [Zn/Fe]$_{\rm exp}\leq -0.3$ and with a small quoted uncertainties (error on the observed [X/Fe] $< 0.1$~dex). These are 11 cases, for all of which  the column density estimates are problematic. 
\begin{enumerate}
\item J0233+0103 was reported to have [Si/Fe]$ = 0.11 \pm 0.07$ \citep{Berg15b}, but the upper limit on $N$(\siii{},1808) is likely underestimated, because it is inconsistent with \siii{} 1304 which is clearly saturated. Therefore [Si/Fe] should be higher than reported.
\item For Q2222-3939, \citet{Berg15} reported [S/Fe] $= 0.00 \pm 0.04$ referring to the measurements of \citet{Noterdaeme08} and \citet{Zafar14b}. However, no information about the \suii{} and \feii{} profiles is given in \citet{Noterdaeme08}. The measurement of \citet{Zafar14b} $\log N(\mbox{\suii{}})=14.25$ seems well estimated, but again no information on Fe is available. From a quick look at the UVES data the \feii{} 1081 and \suii{} 1259 features show a similar strength, which may imply $[S/Fe]\sim+0.3$~dex.
\item J0234-0751 was reported to have [Si/Fe] $= 0.10 \pm 0.09$ \citep{Dutta14}. While \feii{} seems well estimated, the $N$(\siii{},1808) is underestimated due to continuum placement. Therefore the reported [Si/Fe] should be considered as a lower limit.
\item UM673A is the only DLA in this subsample to have a Zn measurement. \citet{Cooke10} reported [Zn/Fe] $= -0.32 \pm 0.00$. The $N$(\znii{}) and $N$(\niii{}) seem abnormally small compared to the other species (for which stronger lines are used in the fit). The measured $N$(\suii{}) suggests that the Zn column may be 0.5 dex larger. The fit of on the \feii{} features may be affected by the extremely small $b$ values assumed in components 1 and 3. Finally, uncertainties in $N$(\znii{}) and $N$(\niii{}) are highly underestimated.
\item Q2344+12 was reported to have [Si/Fe] $= 0.11 \pm 0.03$ \citep{Prochaska01,Prochaska02b}, from high-resolution spectra. While the \siii{},1304 estimates seem reasonable, all \feii{} lines are in the Ly-$\alpha$ forest and likely blended. Hence, the reported [Si/Fe] should be considered as a lower limit.
\item J1241+4617 was reported to have [Si/Fe] $= -0.03 \pm 0.04$ \citep{Rafelski12}, but from low resolution spectra and the absorption lines show a narrow profile. \siii{} 1526 and 1304 are saturated, and therefore the reported column densities should be considered as lower limits. \feii{} 1608 may also be saturated and blended. Therefore the reported [Si/Fe] should be most likely be considered a lower limit.
\item For J1558-0031, \citet{Berg15} reported that [Si/Fe] $= 0.09 \pm 0.04$ referring to the analysis of \citet{Henry07} for high resolution spectra. However, there is no sufficient information on the \feii{} 1608 line profile. The \siii{} 1304 absorption in \citet{OMeara06} is likely saturated, and the quoted error (0.01 dex) seems underestimated, because it highly depends on the $b$ value.
\item J1201+2117 ($z=4.1578$) was reported to have [Si/Fe] $= -0.07 \pm 0.04$ \citep{Rafelski12} from high-resolution spectra. The \siii{} 1526 transition is composed of two narrow components, and may be saturated, and therefore we expect a lower limit for $N$(\siii{}). \feii{} 1608 seems blended in the blue part of the profile, and therefore we expect an upper limit on $N$(\feii{}). Thus, the quoted [Si/Fe] should be considered as a lower limit.
\item J1042+3107 reported a [Si/Fe] $= 0.05 \pm 0.03$ \citep{Rafelski12}. However, low-resolution spectra are used to estimate the column densities, and \siii{} 1526 may be saturated. Therefore the quoted [Si/Fe] should be considered as a lower limit.
\item J1607+1604 reported a [Si/Fe] $= 0.03 \pm 0.06$ from high-resolution spectra. While $N$(\siii{}, 1304) seems  well estimated, \feii{} 1608 is likely blended. Therefore [Si/Fe] should be considered as a lower limit.
\item J0831+4046 reported a [Si/Fe] $= 0.07 \pm 0.08$ \citep{Rafelski12}. However, low-resolution spectra are used to estimate the column densities, and \siii{} 1526 may be saturated. In addition, \feii{} 1608 may be blended. Therefore the quoted [Si/Fe] should be considered as a lower limit. The uncertainties are likely underestimated.
 \end{enumerate}

 \newpage
 
\section{Tables of dust-corrected metallicities}

\begin{table*}[!h]
\centering
\caption{Dust-corrected metallicities and depletions for the clean sample. Additional metallicities and depletions from the DLA sample of \citet{Ledoux06} are already reported in Paper I. References: [a] \citet{Ellison12}; [b] \citet{Peroux06}; [c] \citet{Pettini00}; [d] \citet{Peroux08}; [e] \citet{Rao05}; [f] \citet{Meiring11}; [g] \citet{Pettini99}; [h] \citet{Moller13}; [i] \citet{Rafelski12}; [j] \citet{Ma15}; [k] \citet{Fynbo17}; [l] \citet{Noterdaeme17};  [m] \citet{Noterdaeme10}. $^{aa}$ The sum of $N$(\hi) and 2 $N$(H$_2$) for this strong molecular system.}
\begin{tabular}{ l | r r r r r r r r r r r}
\hline \hline
\rule[-0.2cm]{0mm}{0.8cm}
ID & $z$ & $N$(\hi)  & [Fe/H]$_{\rm tot}$ & [Zn/Fe]$_{\rm exp}$ & $\delta_{\rm Zn}$ & $\delta_{\rm Si}$ & $\delta_{\rm Fe}$ & [$X$/Fe] & $X$ & Ref. \\  
\hline
      B0105-008 & $1.371$ & $21.70\pm 0.15$ & $-1.35\pm0.17$ & $ 0.18$ & $-0.05$ & $-0.15$ & $-0.24$ & $ 0.18\pm 0.05 $ &  Zn   &   a,h  \\
      B2355-106 & $1.172$ & $21.00\pm 0.10$ & $-0.72\pm0.23$ & $ 0.53$ & $-0.14$ & $-0.36$ & $-0.67$ & $ 0.53\pm 0.20 $ &  Zn   &   a,h  \\
     J0000+0048 & $2.525$ & $20.95\pm 0.10^{aa}$ & $ 1.02\pm0.47$ & $ 1.79$ & $-0.49$ & $-1.15$ & $-2.27$ & $ 1.79\pm 0.45 $ &  Zn   &    l  \\
     J0256+0110 & $0.725$ & $20.70\pm 0.20$ & $ 0.13\pm0.37$ & $ 0.90$ & $-0.24$ & $-0.60$ & $-1.15$ & $ 0.90\pm 0.30 $ &  Zn   &   b,h  \\
     J0817+1351 & $4.258$ & $21.30\pm 0.15$ & $-1.19\pm0.23$ & $-0.08$ & $ 0.00$ & $ 0.00$ & $ 0.00$ & $ 0.19\pm 0.22 $ &   S   &  i,h  \\
     J1009+0713 & $0.114$ & $20.68\pm 0.10$ & $-0.71\pm0.24$ & $ 0.06$ & $-0.02$ & $-0.07$ & $-0.08$ & $ 0.30\pm 0.21 $ &   S   &   f,h  \\
     J1051+3107 & $4.139$ & $20.70\pm 0.20$ & $-1.98\pm0.30$ & $ 0.05$ & $-0.01$ & $-0.06$ & $-0.07$ & $ 0.29\pm 0.29 $ &   S   &  i,h  \\
     J1107+0048 & $0.740$ & $21.00\pm 0.04$ & $-0.45\pm0.17$ & $ 0.38$ & $-0.10$ & $-0.26$ & $-0.48$ & $ 0.38\pm 0.15 $ &  Zn   &   b,h  \\
     J1200+4015 & $3.220$ & $20.85\pm 0.10$ & $-0.56\pm0.16$ & $ 0.28$ & $-0.08$ & $-0.21$ & $-0.36$ & $ 0.47\pm 0.15 $ &   S   &  i,h  \\
     J1211+0833 & $2.117$ & $21.00\pm 0.20$ & $ 0.42\pm0.22$ & $ 1.65$ & $-0.45$ & $-1.06$ & $-2.10$ & $ 1.65\pm 0.07 $ &  Zn   &    j  \\
     J1237+0647 & $2.690$ & $20.00\pm 0.15$ & $ 0.77\pm0.17$ & $ 1.30$ & $-0.35$ & $-0.84$ & $-1.64$ & $ 1.30\pm 0.02 $ &  Zn   &    m  \\
     J1431+3952 & $0.602$ & $21.20\pm 0.10$ & $-0.59\pm0.25$ & $ 0.72$ & $-0.20$ & $-0.48$ & $-0.92$ & $ 0.72\pm 0.22 $ &  Zn   &   a,h  \\
     J1438+4314 & $4.399$ & $20.89\pm 0.15$ & $-0.97\pm0.22$ & $ 0.67$ & $-0.18$ & $-0.45$ & $-0.86$ & $ 0.77\pm 0.21 $ &   S   &  i,h  \\
     J1541+3153 & $2.444$ & $20.95\pm 0.10$ & $-1.06\pm0.20$ & $ 0.62$ & $-0.17$ & $-0.42$ & $-0.79$ & $ 0.57\pm 0.19 $ &  Si   &  i,h  \\
     J1607+1604 & $4.474$ & $20.30\pm 0.15$ & $-1.54\pm0.23$ & $-0.19$ & $ 0.00$ & $ 0.00$ & $ 0.00$ & $ 0.17\pm 0.22 $ &  Si   &  i,h  \\
     J1623+0718 & $1.336$ & $21.35\pm 0.10$ & $-0.93\pm0.16$ & $ 0.47$ & $-0.13$ & $-0.33$ & $-0.61$ & $ 0.47\pm 0.10 $ &  Zn   &   a,h  \\
     J2328+0022 & $0.652$ & $20.32\pm 0.06$ & $-0.39\pm0.18$ & $ 0.43$ & $-0.12$ & $-0.30$ & $-0.55$ & $ 0.43\pm 0.15 $ &  Zn   &   b,h  \\
      Q0302-223 & $1.009$ & $20.36\pm 0.11$ & $-0.36\pm0.15$ & $ 0.62$ & $-0.17$ & $-0.42$ & $-0.79$ & $ 0.62\pm 0.07 $ &  Zn   &   c,h  \\
     Q0449-1645 & $1.007$ & $20.98\pm 0.06$ & $-0.88\pm0.12$ & $ 0.37$ & $-0.10$ & $-0.26$ & $-0.48$ & $ 0.37\pm 0.07 $ &  Zn   &   d,h  \\
      Q0454+039 & $0.860$ & $20.69\pm 0.06$ & $-0.98\pm0.14$ & $ 0.00$ & $-0.00$ & $-0.03$ & $-0.01$ & $ 0.00\pm 0.11 $ &  Zn   &   c,h  \\
      Q0933+733 & $1.479$ & $21.62\pm 0.10$ & $-1.44\pm0.12$ & $ 0.37$ & $-0.10$ & $-0.26$ & $-0.47$ & $ 0.37\pm 0.02 $ &  Zn   &   e,h  \\
      Q0948+433 & $1.233$ & $21.62\pm 0.05$ & $-0.97\pm0.09$ & $ 0.43$ & $-0.12$ & $-0.30$ & $-0.55$ & $ 0.43\pm 0.02 $ &  Zn   &   e,h  \\
      Q1354+258 & $1.420$ & $21.54\pm 0.06$ & $-1.47\pm0.20$ & $ 0.41$ & $-0.11$ & $-0.28$ & $-0.52$ & $ 0.41\pm 0.18 $ &  Zn   &   g,h  \\
  eHAQ0111+0641 & $2.027$ & $21.50\pm 0.30$ & $-0.47\pm0.34$ & $ 0.54$ & $-0.15$ & $-0.37$ & $-0.69$ & $ 0.54\pm 0.14 $ &  Zn   &    k  \\
 \hline\hline
 \end{tabular}
\label{tab met clean}
\end{table*}


\begin{flushleft}
\small
Table C.2.. Dust-corrected metallicities and depletions for the large sample. Additional metallicities and depletions are already reported in Paper I. References: [j] \citet{Ma15}; [k] \citet{Fynbo17}; [l] \citet{Noterdaeme17}; [m] \citet{Noterdaeme10}; [n] \citet{Poudel17}; [o] \citet{Berg15} and references therein. $^{aa}$ The sum of $N$(\hi) and 2 $N$(H$_2$) for this strong molecular system. 
\end{flushleft}

\centering
\begin{supertabular}{ l | r r r r r r r r r r r}
\hline \hline
\rule[-0.2cm]{0mm}{0.8cm}
ID & $z$ & $N$(\hi)  & [Fe/H]$_{\rm tot}$ & [Zn/Fe]$_{\rm exp}$ & $\delta_{\rm Zn}$ & $\delta_{\rm Si}$ & $\delta_{\rm Fe}$ & [$X$/Fe] & $X$  & Ref. \\
\hline
         B0105-008 & $1.371$ & $21.70\pm 0.15$ & $-1.35\pm0.17$ & $ 0.18$ & $-0.05$ & $-0.15$ & $-0.24$ & $ 0.18\pm 0.05 $ &  Zn  &    o  \\
        B1036-2257 & $2.778$ & $20.93\pm 0.05$ & $-1.28\pm0.09$ & $ 0.26$ & $-0.07$ & $-0.19$ & $-0.33$ & $ 0.45\pm 0.02 $ &   S  &    o  \\
         B2314-409 & $1.857$ & $20.90\pm 0.10$ & $-0.92\pm0.19$ & $ 0.28$ & $-0.08$ & $-0.21$ & $-0.37$ & $ 0.28\pm 0.14 $ &  Zn  &    o  \\
         B2355-106 & $1.173$ & $21.00\pm 0.10$ & $-0.72\pm0.23$ & $ 0.52$ & $-0.14$ & $-0.36$ & $-0.67$ & $ 0.52\pm 0.20 $ &  Zn  &    o  \\
         BR0951-04 & $3.857$ & $20.60\pm 0.10$ & $-1.18\pm0.14$ & $ 0.59$ & $-0.16$ & $-0.40$ & $-0.75$ & $ 0.55\pm 0.07 $ &  Si  &    o  \\
       BR1108-0747 & $3.608$ & $20.37\pm 0.07$ & $-1.57\pm0.10$ & $ 0.16$ & $-0.04$ & $-0.13$ & $-0.21$ & $ 0.34\pm 0.01 $ &  Si  &    o  \\
       BR1117-1329 & $3.350$ & $20.84\pm 0.12$ & $-1.13\pm0.15$ & $ 0.27$ & $-0.07$ & $-0.20$ & $-0.35$ & $ 0.27\pm 0.04 $ &  Zn  &    o  \\
       BR1202-0725 & $4.383$ & $20.55\pm 0.03$ & $-1.44\pm0.14$ & $ 0.43$ & $-0.12$ & $-0.30$ & $-0.55$ & $ 0.47\pm 0.12 $ &  Si  &    o  \\
      BRI1013+0035 & $3.104$ & $21.10\pm 0.10$ & $-0.11\pm0.13$ & $ 0.99$ & $-0.27$ & $-0.65$ & $-1.26$ & $ 0.99\pm 0.05 $ &  Zn  &    o  \\
            CTQ418 & $2.429$ & $20.68\pm 0.07$ & $-1.72\pm0.11$ & $ 0.23$ & $-0.06$ & $-0.17$ & $-0.29$ & $ 0.37\pm 0.06 $ &  Si  &    o  \\
            CTQ418 & $2.514$ & $20.50\pm 0.07$ & $-1.38\pm0.11$ & $ 0.31$ & $-0.08$ & $-0.22$ & $-0.39$ & $ 0.41\pm 0.04 $ &  Si  &    o  \\
         FJ0812+32 & $2.626$ & $21.35\pm 0.10$ & $-0.58\pm0.14$ & $ 0.90$ & $-0.24$ & $-0.60$ & $-1.15$ & $ 0.90\pm 0.07 $ &  Zn  &    o  \\
         FJ0812+32 & $2.067$ & $21.00\pm 0.10$ & $-1.37\pm0.13$ & $ 0.16$ & $-0.04$ & $-0.13$ & $-0.21$ & $ 0.16\pm 0.03 $ &  Zn  &    o  \\
       FJ2334-0908 & $3.057$ & $20.48\pm 0.05$ & $-0.81\pm0.09$ & $ 0.58$ & $-0.16$ & $-0.40$ & $-0.74$ & $ 0.58\pm 0.04 $ &  Zn  &    o  \\
       HE0515-4414 & $1.150$ & $20.45\pm 0.15$ & $-0.79\pm0.26$ & $ 0.64$ & $-0.17$ & $-0.43$ & $-0.82$ & $ 0.64\pm 0.20 $ &  Zn  &    o  \\
       HE1104-1805 & $1.662$ & $20.85\pm 0.01$ & $-0.84\pm0.07$ & $ 0.55$ & $-0.15$ & $-0.38$ & $-0.71$ & $ 0.55\pm 0.02 $ &  Zn  &    o  \\
       HE1122-1649 & $0.680$ & $20.45\pm 0.05$ & $-0.24\pm0.15$ & $ 1.04$ & $-0.28$ & $-0.68$ & $-1.32$ & $ 0.77\pm 0.12 $ &  Si  &    o  \\
       HE2243-6031 & $2.330$ & $20.67\pm 0.02$ & $-1.03\pm0.08$ & $ 0.14$ & $-0.04$ & $-0.12$ & $-0.19$ & $ 0.14\pm 0.03 $ &  Zn  &    o  \\
       HS0741+4741 & $3.017$ & $20.48\pm 0.10$ & $-1.68\pm0.12$ & $ 0.00$ & $-0.00$ & $-0.03$ & $-0.01$ & $ 0.26\pm 0.01 $ &  Si  &    o  \\
       HS1132+2243 & $2.783$ & $21.00\pm 0.07$ & $-1.77\pm0.16$ & $ 0.35$ & $-0.09$ & $-0.25$ & $-0.44$ & $ 0.43\pm 0.12 $ &  Si  &    o  \\
        J0000+0048 & $2.525$ & $21.07\pm 0.10^{aa}$ & $ 0.90\pm0.47$ & $ 1.79$ & $-0.49$ & $-1.15$ & $-2.27$ & $ 1.79\pm 0.45 $ &  Zn  &    l  \\
        J0008-0958 & $1.768$ & $20.85\pm 0.15$ & $-0.00\pm0.18$ & $ 0.53$ & $-0.14$ & $-0.36$ & $-0.68$ & $ 0.53\pm 0.07 $ &  Zn  &    o  \\
        J0035-0918 & $2.340$ & $20.55\pm 0.10$ & $-2.50\pm0.14$ & $ 0.00$ & $-0.00$ & $-0.03$ & $-0.01$ & $ 0.26\pm 0.07 $ &  Si  &    o  \\
        J0035-0918 & $2.340$ & $20.55\pm 0.10$ & $-2.31\pm0.15$ & $ 0.27$ & $-0.07$ & $-0.20$ & $-0.34$ & $ 0.39\pm 0.08 $ &  Si  &    o  \\
        J0058+0115 & $2.010$ & $21.10\pm 0.15$ & $-0.60\pm0.18$ & $ 0.61$ & $-0.17$ & $-0.41$ & $-0.78$ & $ 0.61\pm 0.07 $ &  Zn  &    o  \\
        J0142+0023 & $3.348$ & $20.38\pm 0.05$ & $-1.57\pm0.14$ & $ 0.31$ & $-0.08$ & $-0.22$ & $-0.39$ & $ 0.41\pm 0.10 $ &  Si  &    o  \\
        J0211+1241 & $2.595$ & $20.60\pm 0.15$ & $-0.64\pm0.19$ & $ 0.35$ & $-0.09$ & $-0.25$ & $-0.44$ & $ 0.43\pm 0.09 $ &  Si  &    o  \\
        J0233+0103 & $1.785$ & $20.60\pm 0.15$ & $-1.33\pm0.18$ & $-0.30$ & $ 0.00$ & $ 0.00$ & $ 0.00$ & $ 0.11\pm 0.07 $ &  Si  &    o  \\
        J0234-0751 & $2.318$ & $20.90\pm 0.10$ & $-1.91\pm0.15$ & $-0.32$ & $ 0.00$ & $ 0.00$ & $ 0.00$ & $ 0.10\pm 0.09 $ &  Si  &    o  \\
          J0255+00 & $3.915$ & $21.30\pm 0.05$ & $-1.78\pm0.12$ & $ 0.07$ & $-0.02$ & $-0.08$ & $-0.10$ & $ 0.31\pm 0.09 $ &   S  &    o  \\
          J0255+00 & $3.253$ & $20.70\pm 0.10$ & $-0.77\pm0.13$ & $ 0.53$ & $-0.14$ & $-0.36$ & $-0.68$ & $ 0.52\pm 0.04 $ &  Si  &    o  \\
        J0256+0110 & $0.725$ & $20.70\pm 0.16$ & $ 0.13\pm0.35$ & $ 0.90$ & $-0.24$ & $-0.60$ & $-1.15$ & $ 0.90\pm 0.30 $ &  Zn  &    o  \\
        J0307-4945 & $4.466$ & $20.67\pm 0.09$ & $-1.36\pm0.22$ & $ 0.35$ & $-0.09$ & $-0.25$ & $-0.44$ & $ 0.43\pm 0.18 $ &  Si  &    o  \\
        J0817+1351 & $4.258$ & $21.30\pm 0.15$ & $-1.22\pm0.18$ & $-0.08$ & $ 0.00$ & $ 0.00$ & $ 0.00$ & $ 0.19\pm 0.06 $ &   S  &    o  \\
        J0824+1302 & $4.472$ & $20.65\pm 0.20$ & $-2.17\pm0.24$ & $-0.12$ & $ 0.00$ & $ 0.00$ & $ 0.00$ & $ 0.20\pm 0.11 $ &  Si  &    o  \\
        J0825+5127 & $3.318$ & $20.85\pm 0.10$ & $-1.49\pm0.16$ & $ 0.35$ & $-0.09$ & $-0.25$ & $-0.44$ & $ 0.43\pm 0.10 $ &  Si  &    o  \\
        J0831+4046 & $4.344$ & $20.75\pm 0.15$ & $-2.10\pm0.18$ & $-0.38$ & $ 0.00$ & $ 0.00$ & $ 0.00$ & $ 0.07\pm 0.08 $ &  Si  &    o  \\
        J0834+2140 & $4.390$ & $21.00\pm 0.20$ & $-1.30\pm0.22$ & $ 0.23$ & $-0.06$ & $-0.17$ & $-0.30$ & $ 0.43\pm 0.04 $ &   S  &    o  \\
        J0834+2140 & $4.461$ & $20.30\pm 0.15$ & $-1.81\pm0.18$ & $-0.12$ & $ 0.00$ & $ 0.00$ & $ 0.00$ & $ 0.20\pm 0.08 $ &  Si  &    o  \\
          J0900+42 & $3.246$ & $20.30\pm 0.10$ & $-0.82\pm0.12$ & $ 0.26$ & $-0.07$ & $-0.19$ & $-0.33$ & $ 0.45\pm 0.01 $ &   S  &    o  \\
        J0909+3303 & $3.658$ & $20.55\pm 0.10$ & $-1.11\pm0.15$ & $ 0.33$ & $-0.09$ & $-0.23$ & $-0.42$ & $ 0.42\pm 0.09 $ &  Si  &    o  \\
        J0953-0504 & $4.203$ & $20.55\pm 0.10$ & $-2.51\pm0.23$ & $-0.04$ & $ 0.00$ & $-0.01$ & $ 0.00$ & $ 0.24\pm 0.19 $ &  Si  &    o  \\
        J0958+0145 & $1.928$ & $20.40\pm 0.10$ & $-0.85\pm0.14$ & $ 0.63$ & $-0.17$ & $-0.42$ & $-0.80$ & $ 0.57\pm 0.08 $ &  Si  &    o  \\
        J1004+0018 & $2.685$ & $21.39\pm 0.10$ & $-1.87\pm0.13$ & $ 0.10$ & $-0.03$ & $-0.09$ & $-0.13$ & $ 0.33\pm 0.04 $ &   S  &    o  \\
        J1004+0018 & $2.540$ & $21.30\pm 0.10$ & $-1.44\pm0.12$ & $ 0.06$ & $-0.02$ & $-0.07$ & $-0.08$ & $ 0.30\pm 0.02 $ &   S  &    o  \\
        J1017+6116 & $2.768$ & $20.60\pm 0.10$ & $-2.01\pm0.14$ & $-0.24$ & $ 0.00$ & $ 0.00$ & $ 0.00$ & $ 0.14\pm 0.06 $ &  Si  &    o  \\
        J1024+0600 & $1.895$ & $20.60\pm 0.15$ & $-0.34\pm0.19$ & $ 0.49$ & $-0.13$ & $-0.34$ & $-0.62$ & $ 0.50\pm 0.09 $ &  Si  &    o  \\
        J1037+0139 & $2.705$ & $20.50\pm 0.08$ & $-1.82\pm0.11$ & $ 0.29$ & $-0.08$ & $-0.21$ & $-0.37$ & $ 0.40\pm 0.04 $ &  Si  &    o  \\
        J1042+0628 & $1.943$ & $20.70\pm 0.15$ & $-0.91\pm0.24$ & $ 0.21$ & $-0.06$ & $-0.16$ & $-0.26$ & $ 0.36\pm 0.17 $ &  Si  &    o  \\
        J1042+3107 & $4.087$ & $20.75\pm 0.10$ & $-1.76\pm0.13$ & $-0.43$ & $ 0.00$ & $ 0.00$ & $ 0.00$ & $ 0.05\pm 0.03 $ &  Si  &    o  \\
        J1049-0110 & $1.658$ & $20.35\pm 0.15$ & $ 0.41\pm0.18$ & $ 0.81$ & $-0.22$ & $-0.54$ & $-1.03$ & $ 0.81\pm 0.07 $ &  Zn  &    o  \\
        J1051+3107 & $4.139$ & $20.70\pm 0.20$ & $-1.94\pm0.21$ & $-0.08$ & $ 0.00$ & $ 0.00$ & $ 0.00$ & $ 0.22\pm 0.04 $ &  Si  &    o  \\
        J1051+3545 & $4.350$ & $20.45\pm 0.10$ & $-1.72\pm0.13$ & $ 0.25$ & $-0.07$ & $-0.18$ & $-0.32$ & $ 0.38\pm 0.05 $ &  Si  &    o  \\
        J1056+1208 & $1.610$ & $21.45\pm 0.15$ & $-0.09\pm0.18$ & $ 0.79$ & $-0.21$ & $-0.53$ & $-1.01$ & $ 0.79\pm 0.07 $ &  Zn  &    o  \\
        J1101+0531 & $4.345$ & $21.30\pm 0.10$ & $-0.93\pm0.20$ & $ 0.51$ & $-0.14$ & $-0.35$ & $-0.65$ & $ 0.51\pm 0.15 $ &  Si  &    o  \\
        J1107+0048 & $0.740$ & $21.00\pm 0.05$ & $-0.45\pm0.17$ & $ 0.37$ & $-0.10$ & $-0.26$ & $-0.48$ & $ 0.37\pm 0.15 $ &  Zn  &    o  \\
        J1111+3509 & $4.052$ & $20.80\pm 0.15$ & $-1.87\pm0.18$ & $-0.14$ & $ 0.00$ & $ 0.00$ & $ 0.00$ & $ 0.19\pm 0.07 $ &  Si  &    o  \\
        J1131+6044 & $2.875$ & $20.50\pm 0.15$ & $-1.06\pm0.21$ & $ 0.88$ & $-0.24$ & $-0.58$ & $-1.11$ & $ 0.69\pm 0.13 $ &  Si  &    o  \\
        J1135-0010 & $2.207$ & $22.05\pm 0.10$ & $-0.86\pm0.13$ & $ 0.70$ & $-0.19$ & $-0.47$ & $-0.90$ & $ 0.70\pm 0.04 $ &  Zn  &    o  \\
        J1142+0701 & $1.841$ & $21.50\pm 0.15$ & $-0.65\pm0.18$ & $ 0.66$ & $-0.18$ & $-0.45$ & $-0.85$ & $ 0.66\pm 0.07 $ &  Zn  &    o  \\
        J1155+0530 & $3.326$ & $21.05\pm 0.10$ & $-0.68\pm0.15$ & $ 0.36$ & $-0.10$ & $-0.26$ & $-0.47$ & $ 0.36\pm 0.09 $ &  Zn  &    o  \\
        J1200+4015 & $3.220$ & $20.65\pm 0.15$ & $-0.30\pm0.18$ & $ 0.39$ & $-0.11$ & $-0.28$ & $-0.50$ & $ 0.39\pm 0.07 $ &  Zn  &    o  \\
        J1201+2117 & $4.158$ & $20.60\pm 0.15$ & $-2.01\pm0.17$ & $-0.67$ & $ 0.00$ & $ 0.00$ & $ 0.00$ & $-0.07\pm 0.04 $ &  Si  &    o  \\
        J1201+2117 & $3.797$ & $21.35\pm 0.15$ & $-0.68\pm0.17$ & $ 0.51$ & $-0.14$ & $-0.35$ & $-0.65$ & $ 0.51\pm 0.04 $ &  Si  &    o  \\
        J1208+0010 & $5.082$ & $20.30\pm 0.15$ & $-1.79\pm0.19$ & $ 0.37$ & $-0.10$ & $-0.26$ & $-0.47$ & $ 0.44\pm 0.09 $ &  Si  &    o  \\
        J1211+0833 & $2.117$ & $21.00\pm 0.20$ & $ 0.42\pm0.22$ & $ 1.65$ & $-0.45$ & $-1.06$ & $-2.10$ & $ 1.65\pm 0.07 $ &  Zn  &    j  \\
        J1219+1603 & $3.003$ & $20.35\pm 0.10$ & $-1.78\pm0.22$ & $-0.45$ & $ 0.00$ & $ 0.00$ & $ 0.00$ & $ 0.04\pm 0.18 $ &  Si  &    o  \\
        J1221+4445 & $4.811$ & $20.65\pm 0.20$ & $-1.58\pm0.22$ & $-1.42$ & $ 0.00$ & $ 0.00$ & $ 0.00$ & $-0.44\pm 0.06 $ &  Si  &    o  \\
        J1237+0647 & $2.690$ & $20.00\pm 0.15$ & $ 0.77\pm0.17$ & $ 1.29$ & $-0.35$ & $-0.84$ & $-1.64$ & $ 1.29\pm 0.02 $ &  Zn  &    m  \\
        J1238+3437 & $2.471$ & $20.80\pm 0.10$ & $-2.05\pm0.17$ & $-0.08$ & $ 0.00$ & $ 0.00$ & $ 0.00$ & $ 0.19\pm 0.11 $ &   S  &    o  \\
        J1240+1455 & $3.108$ & $21.30\pm 0.20$ & $-0.71\pm0.23$ & $ 1.14$ & $-0.31$ & $-0.75$ & $-1.45$ & $ 1.14\pm 0.08 $ &  Zn  &    o  \\
        J1241+4617 & $2.667$ & $20.70\pm 0.10$ & $-1.88\pm0.13$ & $-0.59$ & $ 0.00$ & $ 0.00$ & $ 0.00$ & $-0.03\pm 0.04 $ &  Si  &    o  \\
        J1257-0111 & $4.021$ & $20.30\pm 0.10$ & $-1.25\pm0.14$ & $ 0.61$ & $-0.17$ & $-0.41$ & $-0.78$ & $ 0.56\pm 0.07 $ &  Si  &    o  \\
        J1304+1202 & $2.929$ & $20.30\pm 0.15$ & $-1.55\pm0.17$ & $ 0.19$ & $-0.05$ & $-0.15$ & $-0.25$ & $ 0.40\pm 0.05 $ &   S  &    o  \\
        J1304+1202 & $2.913$ & $20.55\pm 0.15$ & $-1.49\pm0.18$ & $ 0.54$ & $-0.15$ & $-0.37$ & $-0.69$ & $ 0.67\pm 0.06 $ &   S  &    o  \\
        J1305+0924 & $2.018$ & $20.40\pm 0.15$ & $-0.23\pm0.22$ & $ 0.49$ & $-0.13$ & $-0.34$ & $-0.62$ & $ 0.50\pm 0.15 $ &  Si  &    o  \\
        J1310+5424 & $1.801$ & $21.45\pm 0.15$ & $-0.29\pm0.18$ & $ 0.77$ & $-0.21$ & $-0.51$ & $-0.98$ & $ 0.77\pm 0.07 $ &  Zn  &    o  \\
        J1340+1106 & $2.796$ & $21.00\pm 0.06$ & $-1.75\pm0.09$ & $ 0.12$ & $-0.03$ & $-0.11$ & $-0.16$ & $ 0.32\pm 0.02 $ &  Si  &    o  \\
        J1356-1101 & $2.967$ & $20.80\pm 0.10$ & $-1.44\pm0.15$ & $ 0.04$ & $-0.01$ & $-0.06$ & $-0.06$ & $ 0.28\pm 0.09 $ &  Si  &    o  \\
        J1358+0349 & $2.853$ & $20.50\pm 0.10$ & $-2.52\pm0.22$ & $-0.02$ & $ 0.00$ & $-0.02$ & $ 0.00$ & $ 0.25\pm 0.18 $ &  Si  &    o  \\
        J1417+4132 & $1.951$ & $21.45\pm 0.25$ & $-0.29\pm0.27$ & $ 0.81$ & $-0.22$ & $-0.54$ & $-1.03$ & $ 0.81\pm 0.07 $ &  Zn  &    o  \\
        J1419+0829 & $3.050$ & $20.40\pm 0.03$ & $-2.02\pm0.08$ & $-0.02$ & $ 0.00$ & $-0.02$ & $ 0.00$ & $ 0.25\pm 0.03 $ &  Si  &    o  \\
        J1431+3952 & $0.602$ & $21.20\pm 0.10$ & $-0.59\pm0.25$ & $ 0.72$ & $-0.20$ & $-0.48$ & $-0.92$ & $ 0.72\pm 0.22 $ &  Zn  &    o  \\
        J1438+4314 & $4.399$ & $20.89\pm 0.15$ & $-1.18\pm0.17$ & $ 0.52$ & $-0.14$ & $-0.35$ & $-0.66$ & $ 0.65\pm 0.01 $ &   S  &    o  \\
        J1454+0941 & $1.788$ & $20.50\pm 0.15$ & $-0.24\pm0.21$ & $ 0.54$ & $-0.15$ & $-0.37$ & $-0.69$ & $ 0.54\pm 0.13 $ &  Zn  &    o  \\
        J1456+0407 & $2.674$ & $20.35\pm 0.10$ & $-2.02\pm0.17$ & $ 0.39$ & $-0.10$ & $-0.27$ & $-0.50$ & $ 0.45\pm 0.12 $ &  Si  &    o  \\
        J1507+4406 & $3.064$ & $20.75\pm 0.10$ & $-1.99\pm0.16$ & $ 0.03$ & $-0.01$ & $-0.05$ & $-0.05$ & $ 0.28\pm 0.10 $ &   S  &    o  \\
        J1509+1113 & $2.028$ & $21.30\pm 0.15$ & $-0.68\pm0.19$ & $ 0.53$ & $-0.14$ & $-0.36$ & $-0.68$ & $ 0.52\pm 0.09 $ &  Si  &    o  \\
        J1541+3153 & $2.444$ & $20.95\pm 0.10$ & $-1.45\pm0.20$ & $ 0.37$ & $-0.10$ & $-0.26$ & $-0.48$ & $ 0.37\pm 0.16 $ &  Zn  &    o  \\
        J1555+4800 & $2.391$ & $21.50\pm 0.15$ & $-0.25\pm0.18$ & $ 0.84$ & $-0.23$ & $-0.55$ & $-1.06$ & $ 0.67\pm 0.07 $ &  Si  &    o  \\
        J1558-0031 & $2.703$ & $20.67\pm 0.05$ & $-1.79\pm0.09$ & $-0.34$ & $ 0.00$ & $ 0.00$ & $ 0.00$ & $ 0.09\pm 0.04 $ &  Si  &    o  \\
        J1604+3951 & $3.163$ & $21.75\pm 0.20$ & $-1.12\pm0.22$ & $ 0.49$ & $-0.13$ & $-0.34$ & $-0.63$ & $ 0.49\pm 0.07 $ &  Zn  &    o  \\
        J1607+1604 & $4.474$ & $20.30\pm 0.15$ & $-1.56\pm0.18$ & $-0.47$ & $ 0.00$ & $ 0.00$ & $ 0.00$ & $ 0.03\pm 0.06 $ &  Si  &    o  \\
        J1623+0718 & $1.336$ & $21.35\pm 0.10$ & $-0.93\pm0.16$ & $ 0.47$ & $-0.13$ & $-0.33$ & $-0.61$ & $ 0.47\pm 0.10 $ &  Zn  &    o  \\
        J1637+2901 & $3.496$ & $20.70\pm 0.10$ & $-2.02\pm0.21$ & $-1.48$ & $ 0.00$ & $ 0.00$ & $ 0.00$ & $-0.47\pm 0.17 $ &  Si  &    o  \\
        J1712+5755 & $2.253$ & $20.60\pm 0.10$ & $-1.17\pm0.14$ & $ 0.27$ & $-0.07$ & $-0.20$ & $-0.34$ & $ 0.39\pm 0.06 $ &  Si  &    o  \\
        J2036-0553 & $2.280$ & $21.20\pm 0.15$ & $-1.63\pm0.20$ & $ 0.12$ & $-0.03$ & $-0.11$ & $-0.16$ & $ 0.32\pm 0.12 $ &  Si  &    o  \\
        J2321+1421 & $2.573$ & $20.70\pm 0.05$ & $-1.76\pm0.10$ & $-0.06$ & $ 0.00$ & $ 0.00$ & $ 0.00$ & $ 0.23\pm 0.05 $ &  Si  &    o  \\
        J2328+0022 & $0.652$ & $20.32\pm 0.07$ & $-0.39\pm0.18$ & $ 0.43$ & $-0.12$ & $-0.30$ & $-0.55$ & $ 0.43\pm 0.15 $ &  Zn  &    o  \\
          J2340-00 & $2.054$ & $20.35\pm 0.15$ & $-0.20\pm0.19$ & $ 0.49$ & $-0.13$ & $-0.34$ & $-0.63$ & $ 0.49\pm 0.09 $ &  Zn  &    o  \\
       PC0953+4749 & $4.243$ & $20.90\pm 0.15$ & $-2.06\pm0.18$ & $ 0.06$ & $-0.02$ & $-0.07$ & $-0.09$ & $ 0.29\pm 0.08 $ &  Si  &    o  \\
        PKS1354-17 & $2.780$ & $20.30\pm 0.15$ & $-1.45\pm0.19$ & $ 0.63$ & $-0.17$ & $-0.42$ & $-0.80$ & $ 0.57\pm 0.10 $ &  Si  &    o  \\
      PSS1253-0228 & $2.783$ & $21.85\pm 0.20$ & $-1.64\pm0.23$ & $ 0.25$ & $-0.07$ & $-0.19$ & $-0.33$ & $ 0.25\pm 0.08 $ &  Zn  &    o  \\
      PSS1443+2724 & $4.224$ & $20.95\pm 0.10$ & $-0.57\pm0.13$ & $ 0.36$ & $-0.10$ & $-0.25$ & $-0.46$ & $ 0.53\pm 0.03 $ &   S  &    o  \\
      PSS1506+5220 & $3.224$ & $20.67\pm 0.07$ & $-2.10\pm0.11$ & $-0.26$ & $ 0.00$ & $ 0.00$ & $ 0.00$ & $ 0.13\pm 0.04 $ &  Si  &    o  \\
      PSS1802+5616 & $3.811$ & $20.35\pm 0.20$ & $-1.88\pm0.25$ & $-0.20$ & $ 0.00$ & $ 0.00$ & $ 0.00$ & $ 0.16\pm 0.14 $ &  Si  &    o  \\
      PSS2323+2758 & $3.684$ & $20.95\pm 0.10$ & $-2.02\pm0.18$ & $ 0.61$ & $-0.17$ & $-0.41$ & $-0.78$ & $ 0.56\pm 0.13 $ &  Si  &    o  \\
       PSSJ0808+52 & $3.114$ & $20.65\pm 0.07$ & $-1.46\pm0.16$ & $ 0.27$ & $-0.07$ & $-0.20$ & $-0.34$ & $ 0.39\pm 0.13 $ &  Si  &    o  \\
       PSSJ0957+33 & $4.178$ & $20.70\pm 0.10$ & $-1.53\pm0.13$ & $ 0.27$ & $-0.07$ & $-0.20$ & $-0.34$ & $ 0.39\pm 0.05 $ &  Si  &    o  \\
       PSSJ0957+33 & $3.279$ & $20.45\pm 0.08$ & $-0.98\pm0.12$ & $ 0.43$ & $-0.12$ & $-0.30$ & $-0.55$ & $ 0.47\pm 0.05 $ &  Si  &    o  \\
       PSSJ1248+31 & $3.696$ & $20.63\pm 0.07$ & $-1.52\pm0.11$ & $ 0.41$ & $-0.11$ & $-0.28$ & $-0.52$ & $ 0.46\pm 0.05 $ &  Si  &    o  \\
     PSSJ2155+1358 & $3.316$ & $20.50\pm 0.15$ & $-0.97\pm0.38$ & $ 0.38$ & $-0.10$ & $-0.27$ & $-0.49$ & $ 0.38\pm 0.35 $ &  Zn  &    o  \\
     PSSJ2344+0342 & $3.220$ & $21.25\pm 0.08$ & $-1.65\pm0.35$ & $ 0.01$ & $-0.00$ & $-0.04$ & $-0.02$ & $ 0.01\pm 0.34 $ &  Zn  &    o  \\
         Q0000-262 & $3.390$ & $21.41\pm 0.08$ & $-2.03\pm0.12$ & $-0.02$ & $ 0.00$ & $-0.02$ & $ 0.00$ & $-0.02\pm 0.06 $ &  Zn  &    o  \\
         Q0010-002 & $2.025$ & $20.95\pm 0.10$ & $-1.24\pm0.14$ & $-0.15$ & $ 0.00$ & $ 0.00$ & $ 0.00$ & $-0.15\pm 0.06 $ &  Zn  &    o  \\
         Q0013-004 & $1.973$ & $20.83\pm 0.05$ & $-0.50\pm0.10$ & $ 0.77$ & $-0.21$ & $-0.51$ & $-0.98$ & $ 0.77\pm 0.06 $ &  Zn  &    o  \\
        Q0027-1836 & $2.402$ & $21.75\pm 0.10$ & $-1.41\pm0.13$ & $ 0.66$ & $-0.18$ & $-0.45$ & $-0.85$ & $ 0.66\pm 0.03 $ &  Zn  &    o  \\
        Q0039-3354 & $2.224$ & $20.60\pm 0.10$ & $-1.21\pm0.13$ & $ 0.29$ & $-0.08$ & $-0.21$ & $-0.37$ & $ 0.40\pm 0.06 $ &  Si  &    o  \\
        Q0049-2820 & $2.071$ & $20.45\pm 0.10$ & $-1.31\pm0.13$ & $-0.20$ & $ 0.00$ & $ 0.00$ & $ 0.00$ & $ 0.16\pm 0.05 $ &  Si  &    o  \\
         Q0058-292 & $2.671$ & $21.10\pm 0.10$ & $-1.40\pm0.13$ & $ 0.33$ & $-0.09$ & $-0.24$ & $-0.43$ & $ 0.33\pm 0.03 $ &  Zn  &    o  \\
          Q0100+13 & $2.309$ & $21.37\pm 0.08$ & $-1.47\pm0.11$ & $ 0.22$ & $-0.06$ & $-0.17$ & $-0.29$ & $ 0.22\pm 0.01 $ &  Zn  &    o  \\
         Q0102-190 & $2.370$ & $21.00\pm 0.08$ & $-1.88\pm0.11$ & $-0.07$ & $ 0.00$ & $ 0.00$ & $ 0.00$ & $ 0.20\pm 0.03 $ &   S  &    o  \\
         Q0112+030 & $2.423$ & $20.90\pm 0.10$ & $-1.22\pm0.14$ & $ 0.16$ & $-0.04$ & $-0.13$ & $-0.21$ & $ 0.34\pm 0.06 $ &  Si  &    o  \\
         Q0112-306 & $2.418$ & $20.50\pm 0.08$ & $-2.19\pm0.11$ & $-0.16$ & $ 0.00$ & $ 0.00$ & $ 0.00$ & $ 0.18\pm 0.04 $ &  Si  &    o  \\
         Q0112-306 & $2.702$ & $20.30\pm 0.10$ & $-0.41\pm0.14$ & $ 0.55$ & $-0.15$ & $-0.37$ & $-0.70$ & $ 0.53\pm 0.08 $ &  Si  &    o  \\
         Q0135-273 & $2.800$ & $21.00\pm 0.10$ & $-1.52\pm0.20$ & $ 0.00$ & $-0.00$ & $-0.03$ & $-0.01$ & $ 0.26\pm 0.16 $ &  Si  &    o  \\
          Q0149+33 & $2.141$ & $20.50\pm 0.10$ & $-1.59\pm0.16$ & $ 0.14$ & $-0.04$ & $-0.12$ & $-0.19$ & $ 0.14\pm 0.10 $ &  Zn  &    o  \\
        Q0151+0448 & $1.934$ & $20.36\pm 0.10$ & $-1.84\pm0.13$ & $ 0.02$ & $-0.01$ & $-0.04$ & $-0.03$ & $ 0.27\pm 0.05 $ &  Si  &    o  \\
        Q0201+1120 & $3.386$ & $21.26\pm 0.10$ & $-1.27\pm0.19$ & $-0.07$ & $ 0.00$ & $ 0.00$ & $ 0.00$ & $ 0.20\pm 0.14 $ &   S  &    o  \\
          Q0201+36 & $2.463$ & $20.38\pm 0.04$ & $-0.07\pm0.11$ & $ 0.59$ & $-0.16$ & $-0.40$ & $-0.76$ & $ 0.59\pm 0.07 $ &  Zn  &    o  \\
         Q0201+365 & $2.463$ & $20.38\pm 0.05$ & $-0.44\pm0.10$ & $ 0.30$ & $-0.08$ & $-0.22$ & $-0.39$ & $ 0.30\pm 0.05 $ &  Zn  &    o  \\
        Q0216+0803 & $2.293$ & $20.40\pm 0.08$ & $-0.48\pm0.13$ & $ 0.40$ & $-0.11$ & $-0.28$ & $-0.52$ & $ 0.40\pm 0.07 $ &  Zn  &    o  \\
        Q0242-2917 & $2.560$ & $20.90\pm 0.10$ & $-1.86\pm0.13$ & $-0.21$ & $ 0.00$ & $ 0.00$ & $ 0.00$ & $ 0.09\pm 0.04 $ &   S  &    o  \\
        Q0254-4025 & $2.046$ & $20.45\pm 0.08$ & $-1.59\pm0.11$ & $ 0.02$ & $-0.01$ & $-0.04$ & $-0.03$ & $ 0.27\pm 0.04 $ &   S  &    o  \\
        Q0300-3152 & $2.179$ & $20.80\pm 0.10$ & $-1.79\pm0.13$ & $ 0.10$ & $-0.03$ & $-0.09$ & $-0.13$ & $ 0.33\pm 0.04 $ &   S  &    o  \\
         Q0302-223 & $1.009$ & $20.36\pm 0.11$ & $-0.36\pm0.15$ & $ 0.62$ & $-0.17$ & $-0.42$ & $-0.79$ & $ 0.62\pm 0.08 $ &  Zn  &    o  \\
        Q0335-1213 & $3.180$ & $20.78\pm 0.10$ & $-2.19\pm0.20$ & $-0.47$ & $ 0.00$ & $ 0.00$ & $ 0.00$ & $ 0.03\pm 0.16 $ &  Si  &    o  \\
          Q0336-01 & $3.062$ & $21.20\pm 0.10$ & $-1.36\pm0.13$ & $ 0.23$ & $-0.06$ & $-0.17$ & $-0.30$ & $ 0.43\pm 0.03 $ &   S  &    o  \\
          Q0347-38 & $3.025$ & $20.63\pm 0.01$ & $-1.41\pm0.08$ & $ 0.20$ & $-0.06$ & $-0.16$ & $-0.26$ & $ 0.20\pm 0.04 $ &  Zn  &    o  \\
         Q0405-443 & $2.622$ & $20.47\pm 0.10$ & $-1.84\pm0.14$ & $ 0.18$ & $-0.05$ & $-0.14$ & $-0.24$ & $ 0.35\pm 0.06 $ &  Si  &    o  \\
         Q0405-443 & $2.595$ & $21.09\pm 0.10$ & $-0.93\pm0.13$ & $ 0.37$ & $-0.10$ & $-0.26$ & $-0.48$ & $ 0.37\pm 0.03 $ &  Zn  &    o  \\
         Q0405-443 & $2.550$ & $21.13\pm 0.10$ & $-1.22\pm0.14$ & $ 0.33$ & $-0.09$ & $-0.24$ & $-0.43$ & $ 0.33\pm 0.08 $ &  Zn  &    o  \\
        Q0421-2624 & $2.157$ & $20.65\pm 0.10$ & $-1.71\pm0.12$ & $ 0.16$ & $-0.04$ & $-0.13$ & $-0.21$ & $ 0.34\pm 0.01 $ &  Si  &    o  \\
        Q0425-5214 & $2.224$ & $20.30\pm 0.10$ & $-1.37\pm0.13$ & $ 0.26$ & $-0.07$ & $-0.19$ & $-0.33$ & $ 0.45\pm 0.04 $ &   S  &    o  \\
        Q0432-4401 & $2.302$ & $20.95\pm 0.10$ & $-1.18\pm0.17$ & $ 0.23$ & $-0.06$ & $-0.17$ & $-0.29$ & $ 0.37\pm 0.12 $ &  Si  &    o  \\
        Q0449-1645 & $1.007$ & $20.98\pm 0.07$ & $-0.88\pm0.12$ & $ 0.37$ & $-0.10$ & $-0.26$ & $-0.48$ & $ 0.37\pm 0.07 $ &  Zn  &    o  \\
          Q0450-13 & $2.067$ & $20.53\pm 0.08$ & $-1.35\pm0.12$ & $ 0.18$ & $-0.05$ & $-0.14$ & $-0.24$ & $ 0.35\pm 0.04 $ &  Si  &    o  \\
        Q0458-0203 & $2.040$ & $21.65\pm 0.09$ & $-0.98\pm0.13$ & $ 0.59$ & $-0.16$ & $-0.40$ & $-0.76$ & $ 0.59\pm 0.07 $ &  Zn  &    o  \\
        Q0528-2505 & $2.141$ & $20.95\pm 0.05$ & $-1.26\pm0.13$ & $ 0.24$ & $-0.07$ & $-0.18$ & $-0.32$ & $ 0.24\pm 0.09 $ &  Zn  &    o  \\
        Q0528-2505 & $2.812$ & $21.20\pm 0.04$ & $-0.37\pm0.09$ & $ 0.64$ & $-0.17$ & $-0.43$ & $-0.82$ & $ 0.64\pm 0.04 $ &  Zn  &    o  \\
         Q0551-366 & $1.962$ & $20.50\pm 0.08$ & $ 0.13\pm0.13$ & $ 0.81$ & $-0.22$ & $-0.54$ & $-1.03$ & $ 0.81\pm 0.07 $ &  Zn  &    o  \\
        Q0642-5038 & $2.659$ & $20.95\pm 0.08$ & $-1.83\pm0.12$ & $ 0.31$ & $-0.09$ & $-0.23$ & $-0.40$ & $ 0.31\pm 0.05 $ &  Zn  &    o  \\
        Q0824+1302 & $4.809$ & $20.10\pm 0.15$ & $-2.12\pm0.28$ & $-0.24$ & $ 0.00$ & $ 0.00$ & $ 0.00$ & $ 0.14\pm 0.22 $ &  Si  &    n  \\
        Q0824+1302 & $4.829$ & $20.80\pm 0.15$ & $-1.91\pm0.23$ & $ 0.21$ & $-0.06$ & $-0.16$ & $-0.26$ & $ 0.36\pm 0.15 $ &  Si  &    n  \\
          Q0836+11 & $2.465$ & $20.58\pm 0.10$ & $-1.24\pm0.13$ & $ 0.02$ & $-0.01$ & $-0.04$ & $-0.03$ & $ 0.27\pm 0.05 $ &  Si  &    o  \\
          Q0841+12 & $2.476$ & $20.78\pm 0.08$ & $-1.71\pm0.15$ & $ 0.03$ & $-0.01$ & $-0.05$ & $-0.05$ & $ 0.03\pm 0.10 $ &  Zn  &    o  \\
          Q0841+12 & $2.375$ & $20.99\pm 0.08$ & $-1.47\pm0.11$ & $ 0.18$ & $-0.05$ & $-0.15$ & $-0.24$ & $ 0.18\pm 0.02 $ &  Zn  &    o  \\
         Q0913+072 & $2.618$ & $20.34\pm 0.04$ & $-2.38\pm0.08$ & $ 0.02$ & $-0.01$ & $-0.04$ & $-0.03$ & $ 0.27\pm 0.01 $ &  Si  &    o  \\
        Q0918+1636 & $2.412$ & $21.26\pm 0.06$ & $-0.49\pm0.31$ & $ 0.56$ & $-0.15$ & $-0.38$ & $-0.72$ & $ 0.56\pm 0.29 $ &  Zn  &    o  \\
        Q0918+1636 & $2.583$ & $20.96\pm 0.05$ & $ 0.05\pm0.09$ & $ 0.81$ & $-0.22$ & $-0.54$ & $-1.03$ & $ 0.81\pm 0.01 $ &  Zn  &    o  \\
          Q0930+28 & $3.235$ & $20.35\pm 0.10$ & $-2.02\pm0.13$ & $ 0.00$ & $-0.00$ & $-0.03$ & $-0.01$ & $ 0.26\pm 0.04 $ &  Si  &    o  \\
         Q0933+733 & $1.479$ & $21.62\pm 0.10$ & $-1.44\pm0.12$ & $ 0.36$ & $-0.10$ & $-0.26$ & $-0.47$ & $ 0.36\pm 0.02 $ &  Zn  &    o  \\
        Q0933-3319 & $2.682$ & $20.50\pm 0.10$ & $-1.30\pm0.18$ & $ 0.08$ & $-0.02$ & $-0.08$ & $-0.11$ & $ 0.30\pm 0.13 $ &  Si  &    o  \\
         Q0935+417 & $1.373$ & $20.52\pm 0.10$ & $-0.80\pm0.16$ & $ 0.28$ & $-0.08$ & $-0.21$ & $-0.37$ & $ 0.28\pm 0.10 $ &  Zn  &    o  \\
         Q0948+433 & $1.233$ & $21.62\pm 0.06$ & $-0.97\pm0.09$ & $ 0.43$ & $-0.12$ & $-0.30$ & $-0.55$ & $ 0.43\pm 0.01 $ &  Zn  &    o  \\
        Q1010+0003 & $1.265$ & $21.52\pm 0.07$ & $-1.04\pm0.13$ & $ 0.54$ & $-0.15$ & $-0.37$ & $-0.69$ & $ 0.54\pm 0.08 $ &  Zn  &    o  \\
          Q1021+30 & $2.949$ & $20.70\pm 0.10$ & $-1.87\pm0.12$ & $-0.04$ & $ 0.00$ & $-0.01$ & $ 0.00$ & $ 0.24\pm 0.02 $ &  Si  &    o  \\
        Q1036-2257 & $2.778$ & $20.93\pm 0.05$ & $-1.36\pm0.09$ & $ 0.13$ & $-0.03$ & $-0.11$ & $-0.16$ & $ 0.35\pm 0.01 $ &   S  &    o  \\
          Q1055+46 & $3.317$ & $20.34\pm 0.10$ & $-1.64\pm0.17$ & $ 0.02$ & $-0.01$ & $-0.04$ & $-0.03$ & $ 0.27\pm 0.13 $ &  Si  &    o  \\
         Q1111-152 & $3.266$ & $21.30\pm 0.05$ & $-1.51\pm0.13$ & $ 0.35$ & $-0.10$ & $-0.25$ & $-0.45$ & $ 0.35\pm 0.10 $ &  Zn  &    o  \\
        Q1137+3907 & $0.720$ & $21.10\pm 0.10$ & $-0.06\pm0.14$ & $ 0.82$ & $-0.22$ & $-0.55$ & $-1.05$ & $ 0.82\pm 0.07 $ &  Zn  &    o  \\
         Q1157+014 & $1.944$ & $21.70\pm 0.10$ & $-1.09\pm0.14$ & $ 0.46$ & $-0.13$ & $-0.32$ & $-0.59$ & $ 0.46\pm 0.08 $ &  Zn  &    o  \\
         Q1209+093 & $2.584$ & $21.40\pm 0.10$ & $-0.81\pm0.13$ & $ 0.54$ & $-0.15$ & $-0.37$ & $-0.69$ & $ 0.54\pm 0.03 $ &  Zn  &    o  \\
          Q1210+17 & $1.892$ & $20.63\pm 0.08$ & $-0.78\pm0.12$ & $ 0.23$ & $-0.06$ & $-0.18$ & $-0.30$ & $ 0.23\pm 0.06 $ &  Zn  &    o  \\
          Q1215+33 & $1.999$ & $20.95\pm 0.07$ & $-1.13\pm0.12$ & $ 0.42$ & $-0.11$ & $-0.30$ & $-0.54$ & $ 0.42\pm 0.07 $ &  Zn  &    o  \\
        Q1223+1753 & $2.466$ & $21.50\pm 0.10$ & $-1.52\pm0.13$ & $ 0.23$ & $-0.06$ & $-0.18$ & $-0.30$ & $ 0.23\pm 0.04 $ &  Zn  &    o  \\
         Q1232+082 & $2.338$ & $20.90\pm 0.10$ & $-0.57\pm0.18$ & $ 0.87$ & $-0.24$ & $-0.58$ & $-1.11$ & $ 0.87\pm 0.14 $ &  Zn  &    o  \\
         Q1328+307 & $0.692$ & $21.25\pm 0.10$ & $-1.00\pm0.19$ & $ 0.58$ & $-0.16$ & $-0.40$ & $-0.74$ & $ 0.58\pm 0.14 $ &  Zn  &    o  \\
          Q1331+17 & $1.776$ & $21.14\pm 0.08$ & $-1.02\pm0.11$ & $ 0.75$ & $-0.20$ & $-0.50$ & $-0.96$ & $ 0.75\pm 0.04 $ &  Zn  &    o  \\
         Q1337+113 & $2.796$ & $21.00\pm 0.08$ & $-1.73\pm0.11$ & $ 0.14$ & $-0.04$ & $-0.12$ & $-0.19$ & $ 0.33\pm 0.04 $ &  Si  &    o  \\
         Q1354+258 & $1.420$ & $21.54\pm 0.06$ & $-1.46\pm0.13$ & $ 0.42$ & $-0.11$ & $-0.30$ & $-0.54$ & $ 0.42\pm 0.09 $ &  Zn  &    o  \\
        Q1354-1046 & $2.501$ & $20.44\pm 0.05$ & $-1.44\pm0.15$ & $-0.17$ & $ 0.00$ & $ 0.00$ & $ 0.00$ & $ 0.12\pm 0.12 $ &   S  &    o  \\
        Q1354-1046 & $2.967$ & $20.80\pm 0.10$ & $-1.40\pm0.13$ & $ 0.08$ & $-0.02$ & $-0.08$ & $-0.11$ & $ 0.30\pm 0.04 $ &  Si  &    o  \\
         Q1409+095 & $2.456$ & $20.53\pm 0.08$ & $-1.97\pm0.11$ & $-0.02$ & $ 0.00$ & $-0.02$ & $ 0.00$ & $ 0.25\pm 0.03 $ &  Si  &    o  \\
        Q1418-0630 & $3.448$ & $20.40\pm 0.10$ & $-1.21\pm0.15$ & $ 0.31$ & $-0.08$ & $-0.22$ & $-0.39$ & $ 0.41\pm 0.09 $ &  Si  &    o  \\
        Q1425+6039 & $2.827$ & $20.30\pm 0.04$ & $-0.59\pm0.09$ & $ 0.54$ & $-0.15$ & $-0.37$ & $-0.69$ & $ 0.54\pm 0.04 $ &  Zn  &    o  \\
         Q1451+123 & $2.469$ & $20.39\pm 0.10$ & $-1.47\pm0.18$ & $ 0.69$ & $-0.19$ & $-0.46$ & $-0.88$ & $ 0.60\pm 0.14 $ &  Si  &    o  \\
         Q1451+123 & $2.255$ & $20.30\pm 0.15$ & $-0.97\pm0.21$ & $ 0.36$ & $-0.10$ & $-0.26$ & $-0.47$ & $ 0.36\pm 0.13 $ &  Zn  &    o  \\
        Q1502+4837 & $2.570$ & $20.30\pm 0.15$ & $-1.47\pm0.22$ & $-0.43$ & $ 0.00$ & $ 0.00$ & $ 0.00$ & $ 0.05\pm 0.15 $ &  Si  &    o  \\
        Q1727+5302 & $0.945$ & $21.16\pm 0.10$ & $-0.31\pm0.16$ & $ 0.80$ & $-0.22$ & $-0.53$ & $-1.02$ & $ 0.80\pm 0.11 $ &  Zn  &    o  \\
        Q1727+5302 & $1.031$ & $21.41\pm 0.15$ & $-1.06\pm0.29$ & $ 0.79$ & $-0.21$ & $-0.53$ & $-1.01$ & $ 0.79\pm 0.24 $ &  Zn  &    o  \\
         Q1755+578 & $1.971$ & $21.40\pm 0.15$ & $ 0.08\pm0.18$ & $ 0.90$ & $-0.24$ & $-0.60$ & $-1.15$ & $ 0.90\pm 0.07 $ &  Zn  &    o  \\
          Q1759+75 & $2.625$ & $20.76\pm 0.05$ & $-0.79\pm0.09$ & $ 0.31$ & $-0.08$ & $-0.22$ & $-0.39$ & $ 0.41\pm 0.03 $ &  Si  &    o  \\
         Q2059-360 & $3.083$ & $20.98\pm 0.08$ & $-1.71\pm0.14$ & $-0.04$ & $ 0.00$ & $-0.01$ & $ 0.00$ & $ 0.24\pm 0.09 $ &  Si  &    o  \\
         Q2138-444 & $2.852$ & $20.98\pm 0.05$ & $-1.67\pm0.09$ & $ 0.12$ & $-0.03$ & $-0.11$ & $-0.16$ & $ 0.12\pm 0.03 $ &  Zn  &    o  \\
         Q2206-199 & $1.920$ & $20.65\pm 0.07$ & $-0.23\pm0.10$ & $ 0.45$ & $-0.12$ & $-0.31$ & $-0.58$ & $ 0.45\pm 0.02 $ &  Zn  &    o  \\
         Q2206-199 & $2.076$ & $20.43\pm 0.04$ & $-2.16\pm0.08$ & $ 0.04$ & $-0.01$ & $-0.06$ & $-0.06$ & $ 0.28\pm 0.01 $ &  Si  &    o  \\
        Q2222-3939 & $2.154$ & $20.85\pm 0.10$ & $-1.76\pm0.13$ & $-0.33$ & $ 0.00$ & $ 0.00$ & $ 0.00$ & $-0.00\pm 0.04 $ &   S  &    o  \\
          Q2223+20 & $3.119$ & $20.30\pm 0.10$ & $-2.07\pm0.14$ & $ 0.04$ & $-0.01$ & $-0.06$ & $-0.06$ & $ 0.28\pm 0.07 $ &  Si  &    o  \\
        Q2228-3954 & $2.095$ & $21.20\pm 0.10$ & $-1.26\pm0.14$ & $ 0.18$ & $-0.05$ & $-0.15$ & $-0.24$ & $ 0.18\pm 0.06 $ &  Zn  &    o  \\
         Q2230+025 & $1.864$ & $20.83\pm 0.05$ & $-0.52\pm0.11$ & $ 0.45$ & $-0.12$ & $-0.31$ & $-0.58$ & $ 0.45\pm 0.07 $ &  Zn  &    o  \\
          Q2231-00 & $2.066$ & $20.53\pm 0.08$ & $-0.76\pm0.12$ & $ 0.31$ & $-0.09$ & $-0.23$ & $-0.40$ & $ 0.31\pm 0.06 $ &  Zn  &    o  \\
        Q2237-0608 & $4.080$ & $20.52\pm 0.11$ & $-1.78\pm0.18$ & $ 0.06$ & $-0.02$ & $-0.07$ & $-0.09$ & $ 0.29\pm 0.12 $ &  Si  &    o  \\
        Q2311-3721 & $2.182$ & $20.55\pm 0.07$ & $-1.56\pm0.11$ & $ 0.04$ & $-0.01$ & $-0.06$ & $-0.06$ & $ 0.28\pm 0.05 $ &  Si  &    o  \\
        Q2318-1107 & $1.989$ & $20.68\pm 0.05$ & $-0.68\pm0.09$ & $ 0.43$ & $-0.12$ & $-0.30$ & $-0.55$ & $ 0.43\pm 0.03 $ &  Zn  &    o  \\
          Q2342+34 & $2.908$ & $21.10\pm 0.10$ & $-0.67\pm0.15$ & $ 0.84$ & $-0.23$ & $-0.55$ & $-1.06$ & $ 0.67\pm 0.09 $ &  Si  &    o  \\
          Q2343+12 & $2.431$ & $20.34\pm 0.10$ & $-0.60\pm0.13$ & $ 0.42$ & $-0.11$ & $-0.30$ & $-0.54$ & $ 0.42\pm 0.04 $ &  Zn  &    o  \\
          Q2344+12 & $2.538$ & $20.36\pm 0.10$ & $-1.61\pm0.13$ & $-0.30$ & $ 0.00$ & $ 0.00$ & $ 0.00$ & $ 0.11\pm 0.03 $ &  Si  &    o  \\
          Q2348-01 & $2.426$ & $20.50\pm 0.10$ & $-0.33\pm0.12$ & $ 0.94$ & $-0.25$ & $-0.62$ & $-1.19$ & $ 0.72\pm 0.02 $ &  Si  &    o  \\
          Q2348-01 & $2.615$ & $21.30\pm 0.10$ & $-1.87\pm0.17$ & $ 0.04$ & $-0.01$ & $-0.06$ & $-0.06$ & $ 0.28\pm 0.11 $ &  Si  &    o  \\
        Q2348-1444 & $2.279$ & $20.59\pm 0.08$ & $-1.85\pm0.13$ & $ 0.08$ & $-0.02$ & $-0.08$ & $-0.11$ & $ 0.30\pm 0.08 $ &  Si  &    o  \\
          Q2359-02 & $2.154$ & $20.35\pm 0.10$ & $-1.46\pm0.13$ & $ 0.27$ & $-0.07$ & $-0.20$ & $-0.34$ & $ 0.39\pm 0.06 $ &  Si  &    o  \\
          Q2359-02 & $2.095$ & $20.70\pm 0.10$ & $-0.54\pm0.13$ & $ 0.83$ & $-0.23$ & $-0.55$ & $-1.06$ & $ 0.83\pm 0.04 $ &  Zn  &    o  \\
     SDSS0225+0054 & $2.714$ & $21.00\pm 0.15$ & $-0.61\pm0.21$ & $ 0.43$ & $-0.12$ & $-0.30$ & $-0.55$ & $ 0.43\pm 0.14 $ &  Zn  &    o  \\
     SDSS0759+3129 & $3.035$ & $20.60\pm 0.10$ & $-1.97\pm0.38$ & $ 0.00$ & $-0.00$ & $-0.03$ & $-0.01$ & $ 0.26\pm 0.36 $ &  Si  &    o  \\
     SDSS0844+5153 & $2.775$ & $21.45\pm 0.15$ & $-0.70\pm0.18$ & $ 0.77$ & $-0.21$ & $-0.51$ & $-0.98$ & $ 0.64\pm 0.06 $ &  Si  &    o  \\
     SDSS1003+5520 & $2.502$ & $20.35\pm 0.15$ & $-1.27\pm0.46$ & $ 1.22$ & $-0.33$ & $-0.79$ & $-1.55$ & $ 0.86\pm 0.42 $ &  Si  &    o  \\
     SDSS1042+0117 & $2.267$ & $20.75\pm 0.15$ & $-0.90\pm0.23$ & $ 0.18$ & $-0.05$ & $-0.14$ & $-0.24$ & $ 0.35\pm 0.16 $ &  Si  &    o  \\
     SDSS1043+6151 & $2.786$ & $20.60\pm 0.15$ & $-1.82\pm0.40$ & $-0.40$ & $ 0.00$ & $ 0.00$ & $ 0.00$ & $ 0.06\pm 0.36 $ &  Si  &    o  \\
     SDSS1048+3911 & $2.296$ & $20.70\pm 0.10$ & $-2.13\pm0.38$ & $-0.20$ & $ 0.00$ & $ 0.00$ & $ 0.00$ & $ 0.16\pm 0.36 $ &  Si  &    o  \\
    SDSS1116+4118A & $2.662$ & $20.48\pm 0.10$ & $-0.46\pm0.25$ & $ 0.88$ & $-0.24$ & $-0.58$ & $-1.12$ & $ 0.88\pm 0.22 $ &  Zn  &    o  \\
     SDSS1249-0233 & $1.781$ & $21.45\pm 0.15$ & $-0.78\pm0.17$ & $ 0.52$ & $-0.14$ & $-0.36$ & $-0.67$ & $ 0.52\pm 0.05 $ &  Zn  &    o  \\
     SDSS1251+4120 & $2.730$ & $21.10\pm 0.10$ & $-2.05\pm0.44$ & $-1.22$ & $ 0.00$ & $ 0.00$ & $ 0.00$ & $-0.34\pm 0.42 $ &  Si  &    o  \\
     SDSS1435+0420 & $1.656$ & $21.25\pm 0.15$ & $-1.00\pm0.19$ & $-0.16$ & $ 0.00$ & $ 0.00$ & $ 0.00$ & $ 0.18\pm 0.10 $ &  Si  &    o  \\
     SDSS1440+0637 & $2.518$ & $21.00\pm 0.15$ & $-1.74\pm0.46$ & $-1.22$ & $ 0.00$ & $ 0.00$ & $ 0.00$ & $-0.34\pm 0.42 $ &  Si  &    o  \\
     SDSS1557+2320 & $3.538$ & $20.65\pm 0.10$ & $-1.64\pm0.44$ & $ 0.61$ & $-0.17$ & $-0.41$ & $-0.78$ & $ 0.56\pm 0.42 $ &  Si  &    o  \\
     SDSS1610+4724 & $2.508$ & $21.15\pm 0.15$ & $ 0.01\pm0.18$ & $ 0.78$ & $-0.21$ & $-0.52$ & $-1.00$ & $ 0.78\pm 0.07 $ &  Zn  &    o  \\
     SDSS1709+3417 & $2.530$ & $20.45\pm 0.15$ & $-1.47\pm0.33$ & $-0.20$ & $ 0.00$ & $ 0.00$ & $ 0.00$ & $ 0.16\pm 0.28 $ &  Si  &    o  \\
     SDSS2059-0529 & $2.210$ & $20.80\pm 0.20$ & $-0.26\pm0.26$ & $ 0.78$ & $-0.21$ & $-0.52$ & $-1.00$ & $ 0.78\pm 0.16 $ &  Zn  &    o  \\
     SDSS2100-0641 & $3.092$ & $21.05\pm 0.15$ & $-0.23\pm0.18$ & $ 0.71$ & $-0.19$ & $-0.48$ & $-0.91$ & $ 0.71\pm 0.07 $ &  Zn  &    o  \\
     SDSS2222-0946 & $2.354$ & $20.50\pm 0.15$ & $-0.26\pm0.19$ & $ 0.65$ & $-0.18$ & $-0.44$ & $-0.83$ & $ 0.58\pm 0.09 $ &  Si  &    o  \\
    SDSSJ1558+4053 & $2.553$ & $20.30\pm 0.04$ & $-2.31\pm0.10$ & $-0.10$ & $ 0.00$ & $ 0.00$ & $ 0.00$ & $ 0.21\pm 0.06 $ &  Si  &    o  \\
    SDSSJ1616+4154 & $0.321$ & $20.60\pm 0.20$ & $-0.28\pm0.24$ & $ 0.57$ & $-0.15$ & $-0.38$ & $-0.72$ & $ 0.69\pm 0.12 $ &   S  &    o  \\
    SDSSJ1619+3342 & $0.096$ & $20.55\pm 0.10$ & $-0.30\pm0.21$ & $ 1.02$ & $-0.28$ & $-0.67$ & $-1.30$ & $ 1.04\pm 0.17 $ &   S  &    o  \\
            UM673A & $1.626$ & $20.70\pm 0.10$ & $-1.59\pm0.20$ & $-0.32$ & $ 0.00$ & $ 0.00$ & $ 0.00$ & $-0.32\pm 0.15 $ &  Zn  &    o  \\
     eHAQ0111+0641 & $2.027$ & $21.50\pm 0.30$ & $-0.47\pm0.34$ & $ 0.54$ & $-0.15$ & $-0.37$ & $-0.69$ & $ 0.54\pm 0.14 $ &  Zn  &    k  \\
 \hline\hline
 \end{supertabular}
\label{tab met large}

\end{document}